\documentclass[iop]{emulateapj}
\usepackage{graphicx}
\usepackage[flushleft]{threeparttable}
\usepackage{blindtext}
\usepackage{amsmath}
\usepackage{mathtools}
\usepackage{multirow}
\usepackage{hyperref}
\hypersetup{
	colorlinks	= true,
	linkcolor	= red,
	urlcolor	= cyan,
	citecolor	= blue
}

\newcommand{\hst}{\mbox{HST}}
\newcommand{\wfc}{\mbox{WFC3}}
\newcommand{\grism}{\mbox{G141}}

\newcommand{\calwf}{\mbox{CalWF3}}
\newcommand{\axe}{\mbox{aXe}}
\newcommand{\taurex}{\mbox{$\mathcal{T}$-REx}}
\newcommand{\exomol}{\mbox{ExoMol}}
\newcommand{\hitran}{\mbox{HITRAN}}
\newcommand{\hitemp}{\mbox{HITEMP}}

\newcommand{\system}{\mbox{HD\,209458}}
\newcommand{\planet}{\mbox{HD\,209458\,b}}

\begin{document}

% TITLE
\title{A new approach to analysing HST spatial scans: \\ the transmission spectrum of HD\,209458\,b}

\author{A. Tsiaras\altaffilmark{1}, I. P. Waldmann\altaffilmark{1}, M. Rocchetto\altaffilmark{1}, R. Varley\altaffilmark{1}, G. Morello\altaffilmark{1}, M. Damiano\altaffilmark{1,2}, G. Tinetti\altaffilmark{1}}

\affil{$^1$Department of Physics \& Astronomy, University College London, Gower Street, WC1E6BT London, United Kingdom}
\affil{$^2$INAF--Osservatorio Astronomico di Palermo, Piazza del Parlamento 1, I-90134 Palermo, Italy}

\email{angelos.tsiaras.14@ucl.ac.uk}

\begin{abstract}
The Wide Field Camera 3 (\wfc) on Hubble Space Telescope (\hst) is currently one of the most widely used instruments for observing exoplanetary atmospheres, especially with the use of the spatial scanning technique. An increasing number of exoplanets have been studied using this technique as it enables the observation of bright targets without saturating the sensitive detectors. In this work we present a new pipeline for analyzing the data obtained with the spatial scanning technique, starting from the raw data provided by the instrument. In addition to commonly used correction techniques, we take into account the geometric distortions of the instrument, whose impact may become important when combined to the scanning process. Our approach can improve the photometric precision for existing data and also push further the limits of the spatial scanning technique, as it allows the analysis of even longer spatial scans. As an application of our method and pipeline, we present the results from a reanalysis of the spatially scanned transit spectrum of \planet. We calculate the transit depth per wavelength channel with an average relative uncertainty of 40\,ppm. We interpret the final spectrum with \taurex, our fully Bayesian spectral retrieval code, which confirms the presence of water vapor and clouds in the atmosphere of \planet. The narrow wavelength range limits our ability to disentangle the degeneracies between the fitted atmospheric parameters. Additional data over a broader spectral range are needed to address this issue.
\end{abstract}


\maketitle
% TITLE

% INTRODUCTION
\section{INTRODUCTION} \label{sec:introduction}

Transit light-curves have been proved to be an invaluable tool for determining the bulk and orbital parameters of exoplanets. In addition, observations of transits and eclipses at different wavelengths can reveal the thermal structure and composition of the atmosphere. In transmission spectroscopy, atmospheric opacities absorb/scatter small fractions of the stellar light passing through the planetary limb. This imprints a characteristic, wavelength-dependent, variation on the mean transit depth, transmission spectra are 10--100\,ppm in ratio to the radius of the star.
 
Atomic, ionic, molecular and condensate signatures from exoplanetary atmospheres have been identified both with ground-based and space-based instruments \citep[e.g.][]{Charbonneau2002, Vidal2003, Redfield2008, Snellen2008,Swain2009a, Swain2009b, Linsky2010, Swain2010, Tinetti2010, Crouzet2012, Majeau2012, Waldmann2012, Todorov2013, Waldmann2013, Danielski2014, Snellen2014, Sing2016}. 

A recent addition to the capabilities of \hst\ is the spatial scanning technique, which allows the sensitive infrared detector of \wfc\ to observe bright targets. During a spatial scanning exposure the instrument slews slowly along the cross-dispersion direction instead of staring at the target. As a result, the total number of photons collected is much larger, increasing the signal-to-noise ratio (S/N), without the risk of saturation. This observational strategy has already been successfully used to provide an increasing number of exoplanetary spectra \citep[e.g.][]{Deming2013, McCullough2014, Crouzet2014, Fraine2014, Knutson2014a, Knutson2014b, Kreidberg2014b, Kreidberg2014a, Stevenson2014, Kreidberg2015, Tsiaras2016, Line2016}.

The standard \hst\ pipeline, \calwf \footnote{\url{http://www.stsci.edu/hst/wfc3/pipeline/wfc3_pipeline}}, and the spectroscopic package \axe \footnote{\url{http://axe-info.stsci.edu/}} can reduce the \hst\ staring-mode spectroscopic images and extract the 1D spectra from them respectively. By contrast, scanning-mode spectroscopic images have a much more complicated structure that can be described, to the zeroth-order approximation, as the superposition of many staring-mode images, each one slightly shifted along the vertical axis. Due to this structure, only an intermediate product of the \calwf\ package (ima images) is valid when applied to scanning-mode data sets. In addition, the calibration/extraction routines included in the \axe\ package cannot be applied to spatially scanned spectra. In the literature, analyses of data sets obtained in scanning-mode include custom routines to further reduce the ima images and extract the calibrated 1D spectra from them.

In this work we present a stand-alone, dedicated pipeline, able to produce 1D spectra from the raw scanning-mode spectroscopic images. In addition, because of geometrical distortions, the shifted staring-mode spectra that construct each spatially scanned spectrum are not identical to each other (Section \ref{sub:challenges}), a behavior that was either partially or not taken into account in previous analyses. Our pipeline uses a new method to calibrate and extract the 1D spectra (Sections \ref{sub:position} to \ref{sub:extraction}), eliminating possible issues caused by the scanning process. Adopting such an approach allows the efficient analysis of even longer scans, extending the capabilities of the spatial scanning technique.

As an application, we use our new pipeline to reanalyze the \hst/\wfc\ scanning-mode spectroscopic images of the transit of \planet\ \citep{Deming2013}. \planet\ is the very first transiting exoplanet detected \citep{Charbonneau2000} and consequently, the first studied with the transit \citep{Charbonneau2002} and eclipse \citep{Deming2005} spectroscopic methods. Its system parameters can be found in Table \ref{tab:parameters}.

\begin{table}
	\small
	\center
	\caption{\system\ system information}
	\label{tab:parameters}
	\begin{tabular}{l | c }
		
		\hline \hline
		\multicolumn{2}{c}{Stellar parameters}									\\ [0.1ex]
		\hline
		$H$ (mag) $^a$						& 6.591 $\pm$ 0.011				\\
		$J$ (mag) $^a$							& 6.366 $\pm$ 0.035				\\
		$K$ (mag) $^a$						& 6.308 $\pm$ 0.021				\\
		$T_\mathrm{eff}$ (K) $^b$				& 6065 $\pm$ 50				\\
		$[$Fe/H$]$ (dex) $^b$					& 0.00 $\pm$ 0.05				\\
		$M_*$ ($M_{\odot}$) $^b$ 				& 1.119 $\pm$ 0.033				\\
		$R_*$ ($R_{\odot}$) $^b$					& 1.155 $\pm$ 0.016				\\
		$\log(g_*)$ (cgs) $^b$					& 4.361 $\pm$ 0.008				\\ [1.0ex]
		
		\hline \hline
		\multicolumn{2}{c}{Planetary parameters}									\\ [0.1ex]
		\hline
		$T_\mathrm{eq}$ (K) $^b$				& 1449 $\pm$ 12				\\
		$M_\mathrm{p}$ ($M_\mathrm{Jup}$) $^b$	& 0.685 $\pm$ 0.015				\\
		$R_\mathrm{p}$ ($R_\mathrm{Jup}$) $^b$	& 1.359 $\pm$ 0.019				\\
		$a$ (AU) $^b$							& 0.04707 $\pm$ 0.00047			\\

		\hline \hline
		\multicolumn{2}{c}{Transit parameters}									\\ [0.1ex]
		\hline
		$T_0$ (HJD) $^c$					& 2452826.628521 $\pm$ 0.000087		\\
		Period (days) $^c$					& 3.52474859 $\pm$ 0.00000038		\\
		Depth $^{b*}$					 	& 0.014607 $\pm$ 0.000024 			\\
		$T_{14}$ (min) $^{b*}$				& 183.9 $\pm$ 1.1 					\\
		$b$ $^b$							& 0.5070 $\pm$ 005					\\
		$R\mathrm{p}/R_*$ $^b$				& 0.12086 $\pm$ 0.00010				\\
		$a/R_*$ $^b$						& 8.76 $\pm$ 0.04					\\
		$i$ (deg) $^b$						& 86.71 $\pm$ 0.05					\\ [1.0ex]
		
		\multicolumn{2}{l}{$^a$ \cite{Cutri2003}} 									\\
		\multicolumn{2}{l}{$^b$ \cite{Torres2008}, ($^{b*}$ derived)}					\\
		\multicolumn{2}{l}{$^c$ \cite{Knutson2007}}	
		
	\end{tabular}
\end{table}

In terms of composition, transit measurements from space and ground have confirmed the presence of sodium in the atmosphere of \planet\ \citep{Charbonneau2002, Snellen2008, Sing2008}. Other UV observations suggested that the planetary atmosphere is in hydrodynamic escape \citep[e.g.][]{Vidal2003, Holmstrom2008, BenJaffel2010, Linsky2010}. At longer wavelengths, where molecular signatures are stronger, water vapour has been identified by a number of measurements and teams \citep[e.g.][]{Barman2007, Beaulieu2010, Deming2013}. Carbon monoxide has also been detected \citep{Snellen2010} while more carbon species, such as methane and carbon dioxide, have been suggested \citep{Swain2009b, Madhusudhan2009, Line2016}. In addition, the thermal properties of the planet have been investigated by a number of teams \citep[e.g.][]{Burrows2007, Knutson2008, Griffith2014, Line2014, Zellem2014, Schwarz2015, Evans2015}.

In this paper, we use a range of different methods to de-trend the extracted light-curves from the instrumental systematics (Sections \ref{sub:white_lc} and \ref{sub:spectral}) and calculate the transit depth as a function of wavelength. The final spectrum is modeled using the fully Bayesian retrieval framework \taurex\ described in \cite{Waldmann2015b, Waldmann2015a}, based on the Tau code by \cite{Hollis2013}, and using custom generated cross sections based on the line lists from \exomol\ \citep{exomol2014}, \hitran\ \citep{HITRAN2009, HITRAN2013} and \hitemp\ \citep{HITEMP2010} (Section \ref{sec:retrieval}).
% INTRODUCTION

% DATA ANALYSIS
\section{DATA ANALYSIS} \label{sec:analysis}

% Observation - Raw data reduction
\subsection{Observations - Raw data reduction} \label{sub:reduction}

For our analysis we downloaded the spatially scanned spectroscopic images of \planet\ (ID: 12181, PI: Drake Deming) from the MAST Archive \footnote{\url{https://archive.stsci.edu/}}. More specifically, these images are the result of a single visit of the target (containing six \hst\ orbits) using the infrared (IR) detector, the \grism\ grism and a scan rate of 0$''$.9\,s$^{-1}$. Each image consists of 5 non-destructive reads with a size of 266\,$\times$\,266 pixels in the SPARS10 mode, resulting in a total exposure time of 22.32\,s, a maximum signal level of 4.8\,$\times$\,10$^4$\,e$^-$ per pixel and a total scan length of about 170 pixels (0$''$.121091 per pixel). In addition, the data set contains, for calibration purposes, an undispersed (direct) image of the target with the F139M filter.

Our reduction process begins with the raw images, which have not been processed by the standard \hst\ pipeline, \calwf. For this reason we have to apply the basic reduction steps explained in the \textit{WFC3 Data Handbook} \citep[][pp. 55-62]{datahandbook} and the \textit{WFC3 IR Grism Data Reduction Cookbook}\footnote{\url{http://www.stsci.edu/hst/wfc3/documents/WFC3_aXe_cookbook.pdf}} (pp. 16--17). These steps are listed below, and the corresponding \calwf\ routines are stated in brackets. Compared to \calwf, we have modified only the routines that are not suitable for spatially scanned spectroscopic images and also have added the sky background subtraction.

\paragraph{Bias-level and zero-read corrections \\ (ZSIGCORR - BLEVCORR - ZOFFCORR)}

These initial steps are necessary due to the nature of the images, which consist of a number of non-destructive reads, also known as up-the-ramp samples. Our routine follows the implementation of \calwf, beginning with calculating the zero-read flux ($f_\mathrm{z}$). The WFC3 detector lacks a shutter and, as a result, the pixels are collecting photons before the exposure starts. The first non-destructive read of the detector is a reference for all the consecutive ones and it is referred to as the zero-read. $f_\mathrm{z}$ is the illumination recorded in the zero-read, and is important for the non-linearity correction described later. It is calculated by subtracting from the zero-read the super-zero-read frame included in the \textit{u1k1727mi\textunderscore lin.fits} calibration file \citep{linearity}, and stored in memory.

After the calculation of the zero-read flux, the value of reference pixels, located at the beginning and end of each row, are subtracted from each non-destructive read. The reference pixels are not sensitive to incoming light and subtracting them eliminates the 1/f noise between the non-destructive reads. Finally, the zero-read is subtracted from all the consecutive non-destructive reads, as it is the reference level.

\paragraph{Non-linearity correction (NLINCORR)}

The IR detector of the \wfc\ camera is known to perform non-linearly with flux, following the equation:
\begin{equation}
	F_\mathrm{c}(f) = ( 1 + c_1 + c_2 f + c_3 f^2 + c_4 f^3)f
	\label{non_linearity}
\end{equation}

where $F_c$ the collected flux, $f$ is the recorded flux, and $c_n$ are the non-linearity coefficients provided in the \textit{u1k1727mi\textunderscore lin.fits} calibration file. 

This correction is based on the absolute flux in a pixel, and not the difference from the zero-read. Hence, the zero-read flux has to be taken into account. In \calwf\ the amount of flux in the zero-read ($f_\mathrm{z}$) is added to each non-destructive read ($f_\mathrm{r}$) before the correction and subtracted after, so that: $F_\mathrm{final} = F_\mathrm{c}(f_r+f_z) - f_z$. In scanning-mode images, pixels with large zero-read fluxes are very common (very bright targets), and for those pixels $F_\mathrm{final}$ is overestimated. To avoid this additional flux we also correct $f_\mathrm{z}$ before subtracting, so that: $F_\mathrm{final} = F_\mathrm{c}(f_r+f_z) - F_\mathrm{c}(f_z)$. For pixels where $f_\mathrm{z}$ is close to the saturation limit (70,000\,e$^-$), the difference is of the order of 1,000\,e$^-$.

\paragraph{Dark current subtraction (DARKCORR)}

The dark current in the \wfc/IR detector is non-linear with time and also depends on the sub-array mode and the sampling process. In agreement with \calwf, we select from the provided super-dark files \citep{dark} the one that matches with the data set and subtract the respective dark current frame from each non-destructive read.

\paragraph{Gain variations correction (FLATCORR)}

At this step DN units are converted to electrons, in the same way as in \calwf, while taking into account the gain variations between the four quadrants. Each non-destructive read is divided by the pixel flat-field frame included in the \textit{u4m1335mi\textunderscore pfl.fits} calibration file, and multiplied by the mean gain of the four amplifiers (mean gain = 2.35\,e$^-$DN$^{-1}$).

\paragraph{Sky background subtraction}

The sky background subtraction is not included in \calwf. According to the \textit{WFC3 IR Grism Data Reduction Cookbook} (pp. 16--17), the master-sky frame included in the \textit{WFC3.IR.G141.sky.V1.0.fits} calibration file \citep{sky} has to be scaled and subtracted from the images prior to applying the wavelength-dependent flat-field (Section \ref{sub:wavelength}). This is a relative sky background template, which takes into account the variations of the sky background across the detector. The scaling factor is calculated from dividing the least illuminated area of the image by the master-sky frame. For the case of \planet, we use an area on the left side of the spectrum. We avoid both the top and the bottom of the image due to the extended wings of the spectrum and a staring-like ``ghost'' spectrum, respectively. The later is, possibly, the result of persistence from previous observations.

\paragraph{Bad pixels and cosmic rays correction (CRCORR)}

The final step in our reduction process is the correction of bad pixels and cosmic rays. Bad pixels have been identified during the calibration cycles and stored in the calibration file \textit{y711520di\textunderscore bpx.fits} \citep{badpix}. On the contrary, cosmic rays are randomly positioned on the detector and have to be identified in each image, independently. The cosmic rays detection and correction routine included in \calwf\ is based on the assumption that the flux in each pixel increases linearly with time. This behavior is expected for a static source but not for a moving one and, consequently, the above assumption is not valid for scanning-mode data sets. To identify cosmic rays we calculate two flags for each pixel; the difference from the average of the four horizontally neighboring pixels (x-flag) and the difference from the average of the four vertically neighboring pixels (y-flag). If a pixel's x-flag is 5$\sigma$ larger than the other pixels in the column and it's y-flag 5$\sigma$ larger than the other pixels in the row it is identified as a cosmic ray. In this way we take into account the structure of the spatially scanned spectrum along both axes. We correct both the bad pixels (apart from the ``blobs'', that are not single pixels) and the cosmic rays by performing a 2D interpolation of the scientific image excluding those pixels and then filling the gaps with the values of the interpolated function. We have to note here that, in \calwf, the CRCORR step is applied before the FLATCORR step but we choose to apply our routine at the end, to avoid propagating the interpolation uncertainties.

\begin{figure}
	\centering
	\includegraphics[width=\columnwidth]{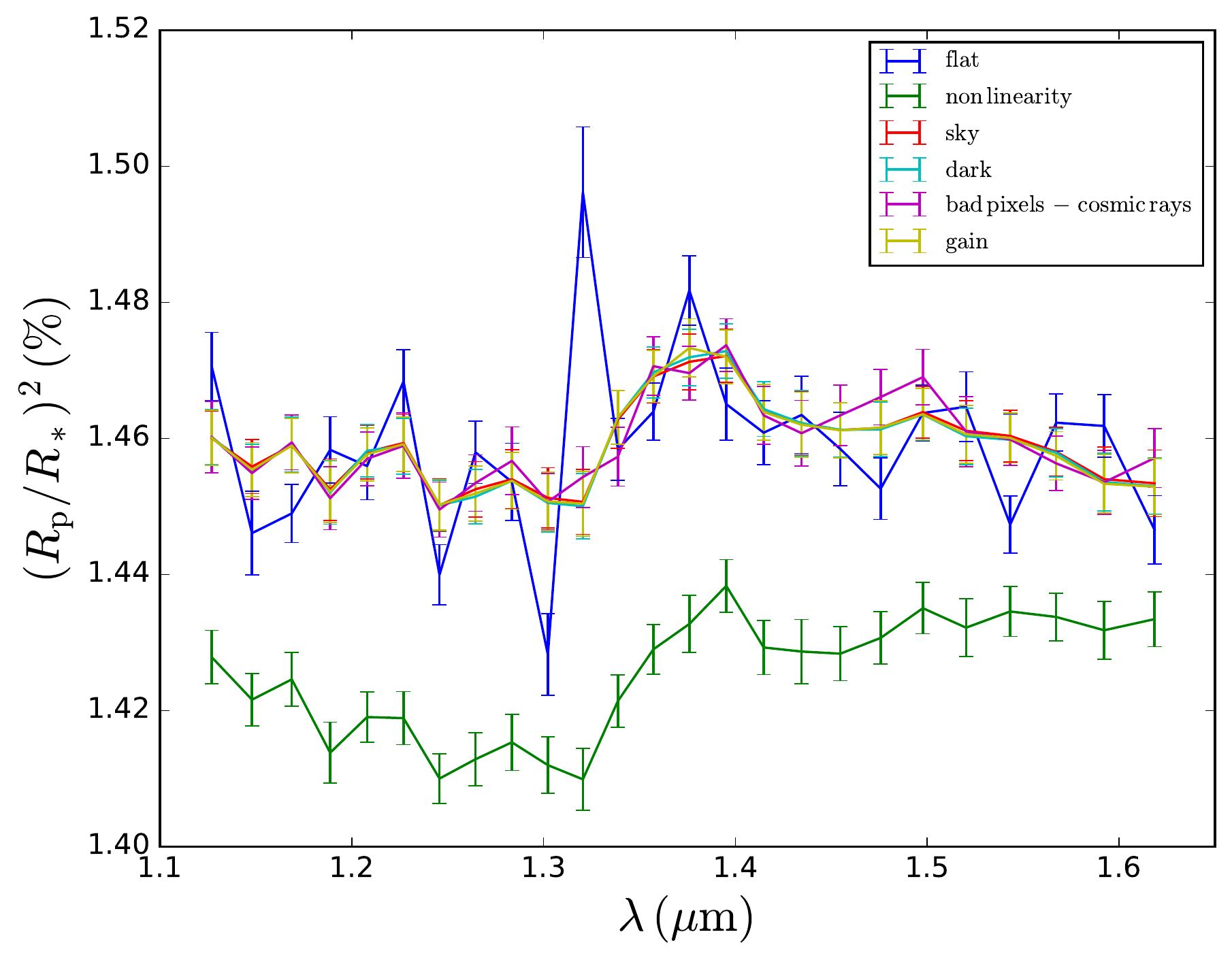}
	\caption{The \planet\ spectrum when switching off each reduction step, as indicated in the legend. For this data set, the reduction steps with the stronger effect are the flat-field (described in Section \ref{sub:flat-field}), the bad pixels/cosmic rays, and the non-linearity corrections. The first two, are introducing strong scatter, while the third one is shifting and distorting the shape of the spectrum.}
	\label{fig:no_corrections}
\end{figure}

In our pipeline, there is the option of omitting any of the above reduction steps, and so, evaluating the effect of each one on the final spectrum. For \planet, we extracted the planetary spectrum (as described in the following sections) for cases where each reduction step is omitted, apart from the initial bias-level and zero-read corrections. Figure \ref{fig:no_corrections} shows the results, where we can see that the flat-field (described in Section \ref{sub:flat-field}), the bad pixels/cosmic rays, and the non-linearity corrections  the have the stronger effects. However, these results are expected to vary between data sets, depending on the specific characteristics of each data set.
% Observation - Raw data reduction

% Structure of the spatially scanned spectra and extraction challenges
\subsection{Structure of the spatially scanned spectra and extraction challenges} \label{sub:challenges}

A spatially scanned spectrum can be described as the superposition of many staring-mode spectra (``building blocks''), each one slightly shifted along the vertical axis of the detector. The most common approach to produce 1D spectra from 2D spatially scanned ones, is to sum along the detector columns. However, the ``building blocks'' of a spatially scanned spectrum are neither identical to each other nor parallel to the detector rows, because:
\begin{enumerate}
	\item there are significant dispersion variations along the vertical axis of the WFC3/IR detector (from about 4.47 to 4.78 nm/pix), caused by the 24 degrees tilt about its horizontal axis,
	\item the $1^\mathrm{st}$ order spectrum of the G141 grism, that is used, is inclined by 0.5 degrees with respect to the WFC3/IR detector rows,
\end{enumerate}
\noindent as described in the \textit{WFC3 Instrument Handbook} \citep[][pp. 173--174]{instrumenthandbook}.

Because of the dispersion variations, the wavelength associated to a detector column is increasing towards its upper part. In the case of \planet\ (scan length of 170 pixels), for a column at 1.2\,$\mu$m, the wavelength difference between the lower and the upper edge of the spatially scanned spectrum is 30\,$\AA$, while at 1.6\,$\mu$m the difference is 70\,$\AA$. These values correspond to 0.6 and 1.5 pixels, respectively. As a result, 1D spectra resulting from summing along the columns of the detector vary by up to 1\% between an intermediate scan of 60 pixels and the final scan of 170 pixels. For longer scans, such as 55 Cancri e \citep[][340 pixels]{Tsiaras2016}, the effect is stronger and the discrepancy can be more that 2\% (Section \ref{sub:extraction}). An effort to correct for dispersion variations has been made by \cite{Kreidberg2014a} with a row-by-row interpolation which rearranges the flux in each row to create a uniformly repeated spectrum along the scanning direction. Although this is a possible approach it may restrict the achievable precision level because the dispersion direction is inclined by 0.5 degrees and, therefore, the ``building blocks'' of the spatially scanning spectrum are not parallel to the detector rows.

Moreover, the inclined spectrum affects the wavelength calibration, as the wavelength solutions depend on the position of a pixel along the trace \textemdash \, i.e. the curve on which the spectrum lies \textemdash \, and not along the x-axis of the detector (see \textit{aXe User Manual version 2.3}\footnote{\url{http://www.stsci.edu/institute/software_hardware/stsdas/axe/extract_calibrate/axe_manual}}, pp. 76-77). The effect of summing along the columns in the wavelength calibration is evident in \cite{Wilkins2014}, where the authors find an inconsistency between the extracted 1D stellar spectrum and the sensitivity curve the G141 grism \citep{coefficients}. The empirical adjustment of the calibration coefficients that these authors propose is up to 10\%.

\hypertarget{w.d.p.t.}{} 

To take into account the effects described above, we follow a calibration process (Sections \ref{sub:position} and \ref{sub:wavelength}) that monitors how the position of the dispersed photons changes during a scan, and define the wavelength-dependent photon trajectories (\hyperlink{w.d.p.t.}{w.d.p.t.}). We then use them to extract 1D spectra that are both consistent with the structure of the spatially scanned spectra, and agree with the sensitivity curve of the G141 grism (Section \ref{sub:extraction}). 
% Structure of the spatially scanned spectra and extraction challenges

% Position shifts
\subsection{Position shifts} \label{sub:position}

\begin{figure}
	\centering
	\includegraphics[width=\columnwidth]{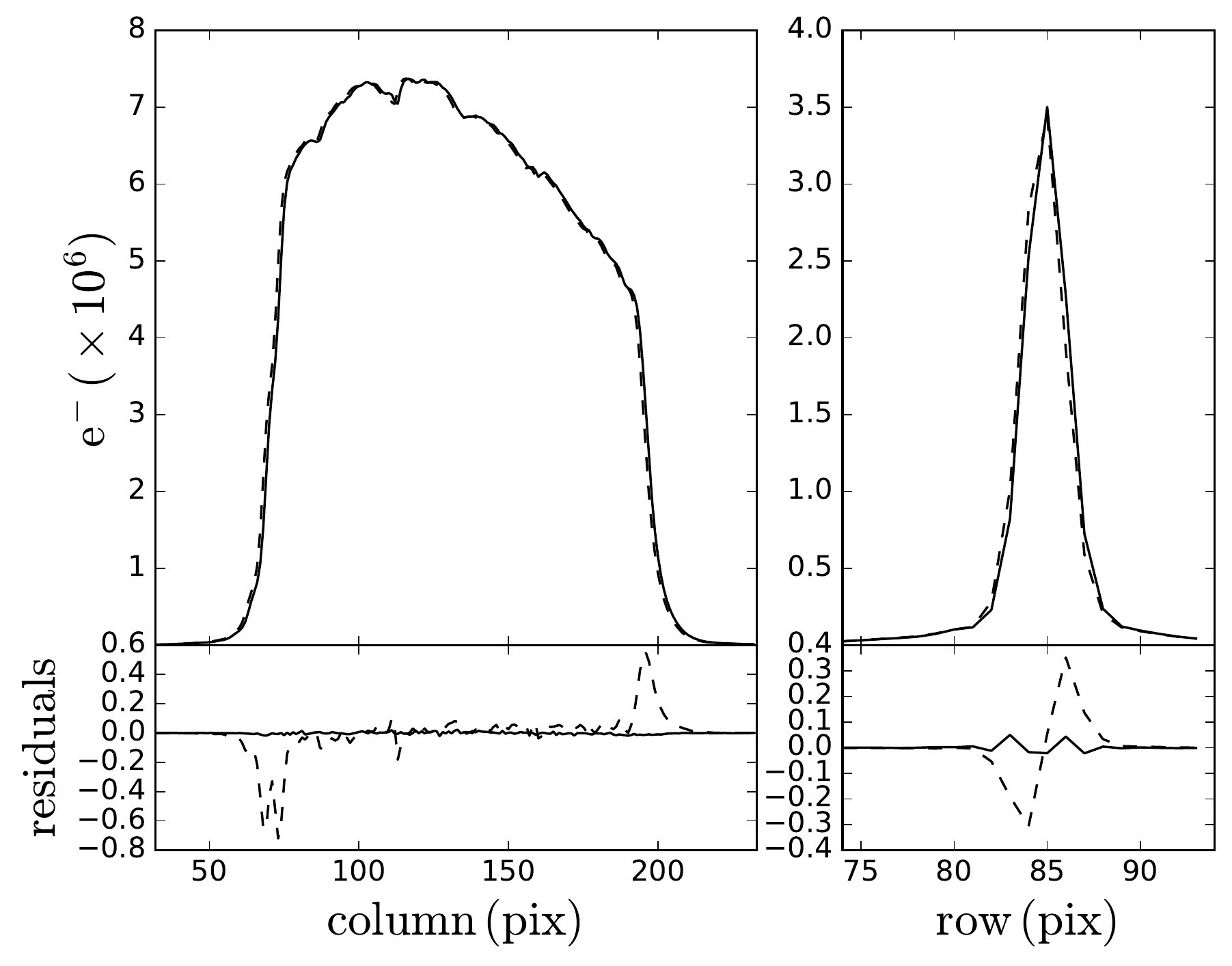}
	\caption{Left-Top: Normalized sum along the columns of the first (continuous) and the last (dashed, normalized) spatially scanned spectra of the visit. Left-Bottom: Difference between the two profiles before and after shifting, dashed and continuous lines, respectively. Right: same plots for the sum along the rows of the first non-destructive read.}
	\label{fig:position}
\end{figure}

While \hst\ guiding system is stable, it fails to reset at exactly the same position as it was before a spatial scanning observation, causing horizontal position shifts of about one pixel over an entire visit \citep{Deming2013, Kreidberg2014a, Knutson2014b, Fraine2014}.

To calculate the horizontal shifts we compare the structure of the first spatially scanned spectrum with all subsequent spectra, using the normalised sum along their columns (Figure \ref{fig:position}, left), similarly to \cite{Kreidberg2014a}. For each consecutive image, i, we interpolate and fit for the horizontal shift, $\Delta x_{i}$, relative to the first one. Note here that the sums used above are corrected for the static (non wavelength-dependent) component of the flat-field, to avoid the bias introduced by its structure. The final values will be used in the following section to define the \hyperlink{w.d.p.t.}{w.d.p.t.}, and therefore the left and right edges of the extraction apertures.

\begin{figure}
	\centering
	\includegraphics[width=\columnwidth]{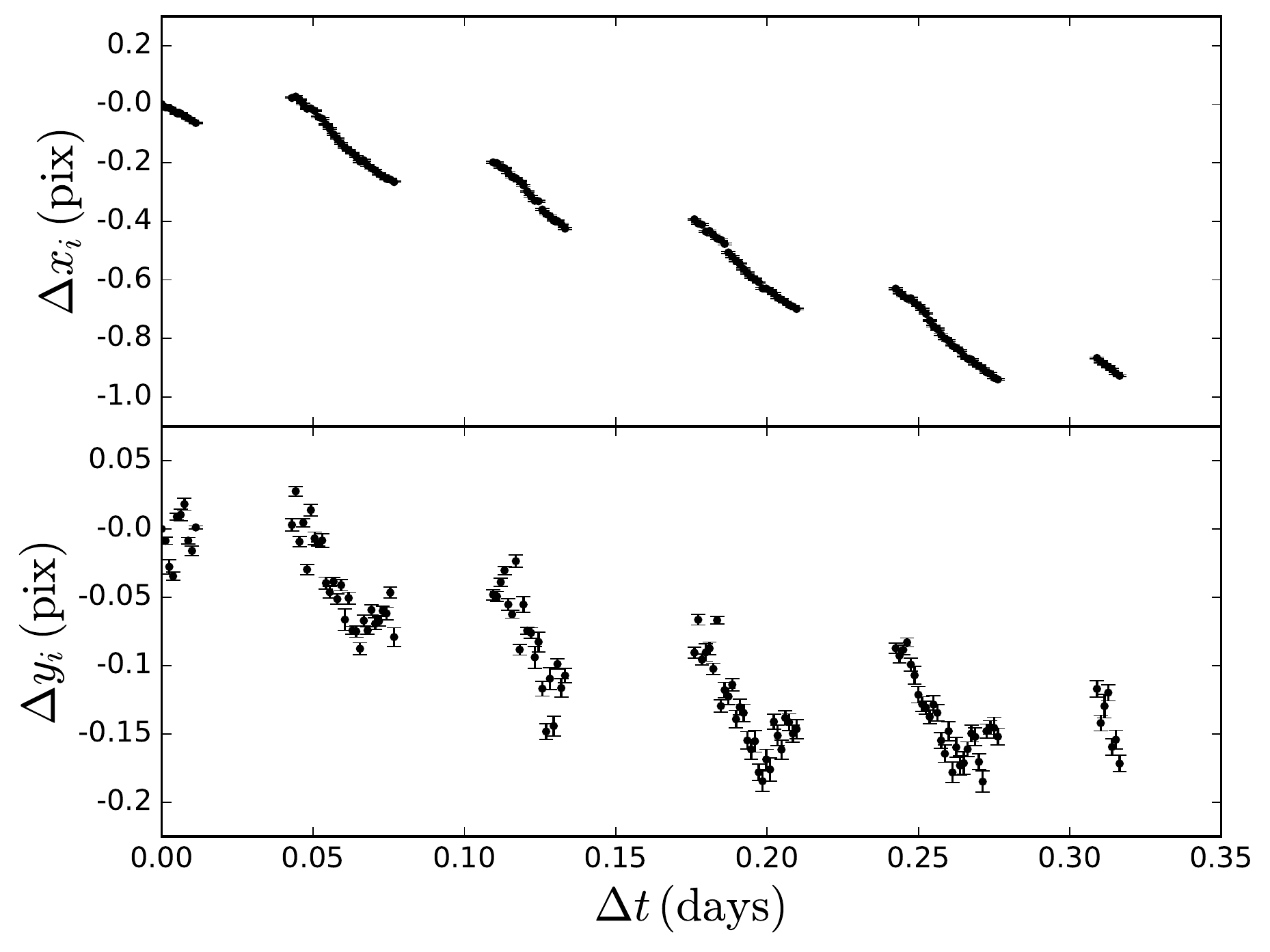}
	\caption{Horizontal (top) and vertical (bottom) shift for each image of the visit, relative to the first one.}
	\label{fig:shifts}
\end{figure}

Horizontal shifts are important as they displace the spectrum on the detector and also introduce additional systematics to the spectral light-curves, such as under-sampling \citep{Deming2013, Wilkins2014}. For this particular data set, we find horizontal shifts of about 0.9 pixel over the visit (top panel in Figure \ref{fig:shifts}). If not taken into account, such shifts introduce variations of up to 250\,ppm in the planetary spectrum.

In addition, shifts of the vertical position from which the scan starts ($\Delta y_{i}$) are calculated from the first non-destructive read of each exposure. We apply the same method as for the horizontal shifts described above, with the difference that here we sum along the rows instead of the columns (Figure \ref{fig:position}, right). Finally, we calculate the scan length ($l_{i}$)  by fitting an extended Gaussian function on the sum along the rows of the last non-destructive read. The results will be used later to define the upper and lower edges of the extraction apertures. For this observation, both vertical shifts (0.2 pixel over the visit, bottom panel in Figure \ref{fig:shifts}) and length variations ($l_{i}$ = 164.688\,$\pm$\,0.017\,pixels over the visit) are not significant enough to affect the final planetary spectrum.

% Position shifts

% Wavelength calibration
\subsection{Wavelength calibration} \label{sub:wavelength}

\subsubsection{Position of the star}

The key information for calibrating a WFC3/G141 spectrum is the physical position of the star ($x^*,y^*$) on the full detector array \citep{coefficients}. For this purpose, every spectrum should be accompanied by an undispersed (direct) image of the star, taken with the filter F140W. In the case of spatially scanned spectra, the vertical position ($y^*$) is not constant and so it cannot be determined from the direct image. In contrast, the horizontal position ($x^*$) is given by the equation:
\begin{equation}
	x^*=	x_0 + (507-0.5L) + \Delta x_\mathrm{off} + \Delta x_\mathrm{ref}
	\label{pos_first}
\end{equation}

\noindent where $x_0$ is the result of fitting a 2D Gaussian function to the direct image, $L$ is the size of the direct image array, $507-0.5L$ is the difference between the coordinate systems of the sub-array used for the direct image and the full detector array (this correction gives the absolute position on the detector, and the number 507 is used because the calibration coefficients do not take into account the reference pixels), $\Delta x_\mathrm{off}$ is the difference in the centroid offsets along the x-axis between the filter used for the direct image and the filter F140W, and $\Delta x_\mathrm{ref}$ is the difference in the chip reference pixels between the WFC3 aperture used for the direct image and the WFC3 aperture used for the dispersed image. Details and tables of values for each one of the above correction parameters can be found in Appendix \ref{app:target}. For \planet, these values are (in pixels): $x_0=137.5$ (for the first scan), $L=256$ (sub-array used: SQ256SUB), $\Delta x_\mathrm{off}=0.027$ (filter used: F139M), $\Delta x_\mathrm{ref}=-107$ (WFC3 apertures used: direct image: IRSUB256, spatial scans: GRISM256)

The limited observational time in each \hst\ visit allows the observers to include only one undispersed image, at the beginning of each visit. With this image we can calculate the $x^*$ that corresponds to the first scan ($x^*_1$), but, for any subsequent scan, we have to use the horizontal shifts calculated in the previous section:
\begin{equation}
	x^*_i = x^*_1 + \Delta x_{i}
	\label{pos_ith}
\end{equation}

\subsubsection{Calculating the wavelength-dependent photon trajectories (w.d.p.t.) }

\begin{figure}
	\centering
	\includegraphics[width=\columnwidth]{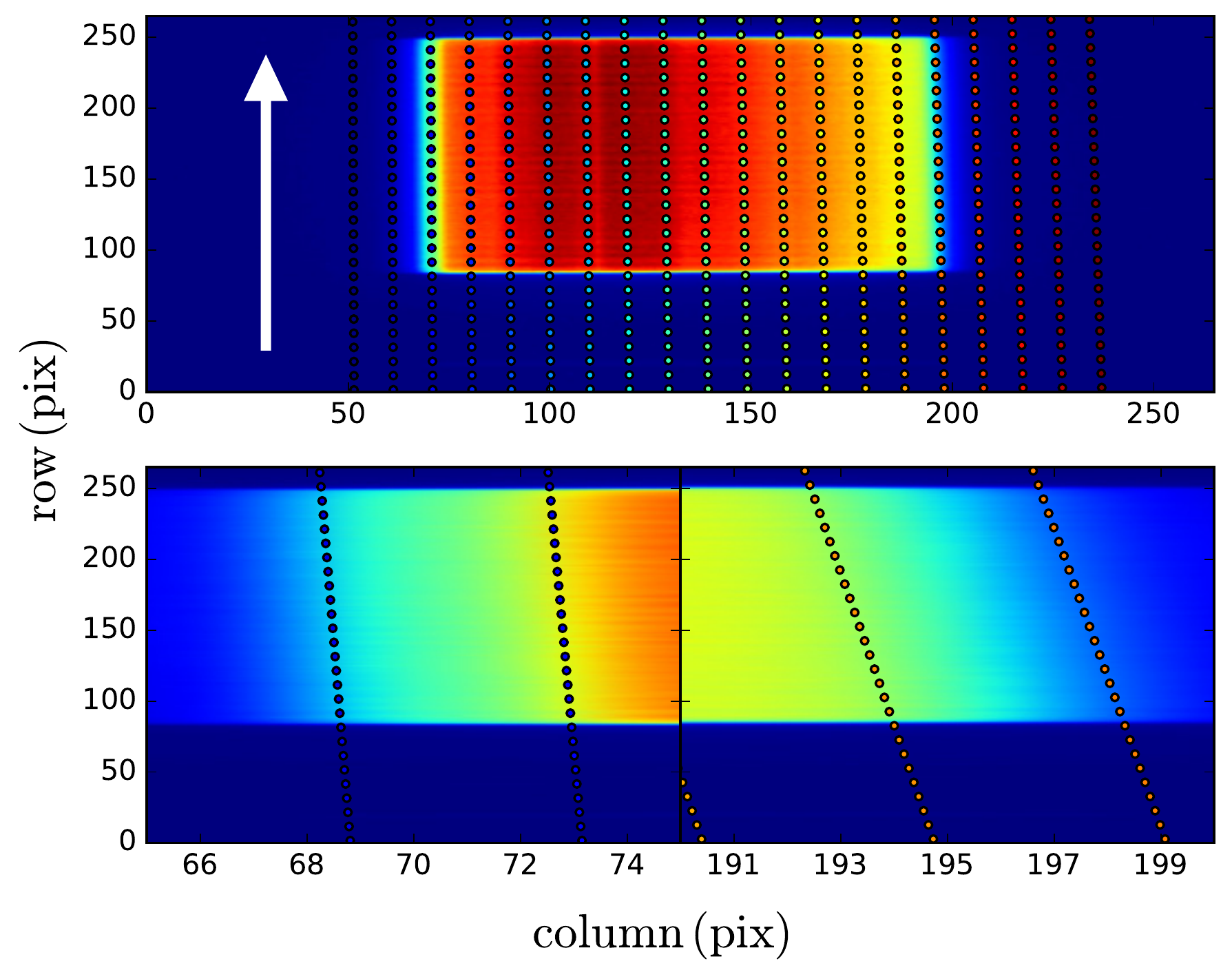}
	\caption{Top: Position of the dispersed photons with different wavelengths (colored points) as the star moves along its scanning trajectory (white arrow). Bottom: Left and right edges of the spectrum where we can appreciate how accurately the grid follows the data.}
	\label{fig:grid}
\end{figure}

As described in the \textit{aXe User Manual version 2.3} (pp. 76-77), the trace of a staring-mode spectrum on the detector is:
\begin{equation}
	y - y^* = a_\mathrm{t}(x - x^*) + b_\mathrm{t} 
	\label{trace}
\end{equation}

\noindent where
\begin{equation*}
	\begin{split}
		\underbrace{a_\mathrm{t}}_{\text{\tiny or DYDX\textunderscore A\textunderscore 1}} =		&\	{\begin{cases} a_\mathrm{t0} + a_\mathrm{t1}x^* + a_\mathrm{t2}y^* + \\ a_\mathrm{t3}x^{*2} + a_\mathrm{t4}x^*y^* + a_\mathrm{t5}y^{*2} \end{cases}} \\
		\underbrace{b_\mathrm{t}}_{\text{\tiny or DYDX\textunderscore A\textunderscore 0}} = 		&\	b_\mathrm{t0} + b_\mathrm{t1}x^* + b_\mathrm{t2}y^*
	\end{split}
\end{equation*}

\noindent and the wavelength solution is:
\begin{equation}
	\lambda =	a_\mathrm{w} d +b_\mathrm{w} 
	\label{wlsolution}
\end{equation}

\noindent where
\begin{equation*}
	\begin{split}
		\underbrace{a_\mathrm{w}}_{\text{\tiny or DLDP\textunderscore A\textunderscore 1}} = 		&\	{\begin{cases} a_\mathrm{w0} + a_\mathrm{w1}x^* + a_\mathrm{w2}y^* + \\ a_\mathrm{w3}x^{*2} + a_\mathrm{w4}x^*y^* + a_\mathrm{w5}y^{*2} \end{cases}} \\
		\underbrace{b_\mathrm{w}}_{\text{\tiny or DLDP\textunderscore A\textunderscore 0}} = 		&\	b_\mathrm{w0} + b_\mathrm{w1}x^* + b_\mathrm{w2}y^*
	\end{split}
\end{equation*}

\noindent and $(x^*,y^*)$ is the physical position of the star on the full detector array, $d$ is the distance from the source along the trace and $(a_{\mathrm{t}n},b_{\mathrm{t}n},a_{\mathrm{w}n},b_{\mathrm{w}n})$ are the \hst\ calibration coefficients included in the configuration file \textit{WFC3.IR.G141.V2.5.conf} \citep[][Appendix \ref{app:grid}]{coefficients}.

In the case of spatially scanned spectra, the star is moving on the detector. We track the changes in the positions of the dispersed photons during each scan and define the \hyperlink{w.d.p.t.}{w.d.p.t.} by following these steps:
\begin{itemize}
	\item work out the position of the dispersed photons on the main trace ($x_{\lambda}, y_{\lambda}$) as function of $y^*$ and wavelength ($\lambda$), using equations \ref{trace} and \ref{wlsolution} (for the proof of the following equations see Appendix \ref{app:grid}) :
\end{itemize}
\begin{equation}
	\begin{split}
		x_{\lambda} = 	&\	x^*- \frac{a_\mathrm{t} b_\mathrm{t} }{1 + a_\mathrm{t}^2} + \frac{\lambda-b_\mathrm{w}}{a_\mathrm{w}} \cos [ \tan^{-1} (a_\mathrm{t}) ] \\
		y_{\lambda} =	&\	a_\mathrm{t}(x_{\lambda} - x^*) + b_\mathrm{t} + y^*
	\end{split}
	\label{wlpos}
\end{equation}

\begin{itemize}
	\item assume $x^*$ to be constant during a scan, but different from one scan to another ($x^*_i$ from Equation \ref{pos_ith}),
	\item let $y^*$ to vary  uniformly across the length of the sub-array, corresponding to the vertical scan,
	\item let $\lambda$ to vary uniformly from 1 to 1.8\,$\AA$, covering the whole response range of the G141 grism,
	\item from all the ($y^*$, $\lambda$) pairs, use equations \ref{wlpos} to create a large grid of ($\lambda, x_{\lambda}, y_{\lambda}$) points (Figure \ref{fig:grid}),
	\item fit on the grid points the function of the \hyperlink{w.d.p.t.}{w.d.p.t.}:
\end{itemize}
\begin{equation}
	y _{\lambda} = (\frac{c_1}{c_2+\lambda} + c_3) + (\frac{s_1}{s_2+\lambda} + s_3) x_{\lambda}
	\label{wlcalib}
\end{equation}

\begin{figure}
	\centering
	\includegraphics[width=\columnwidth]{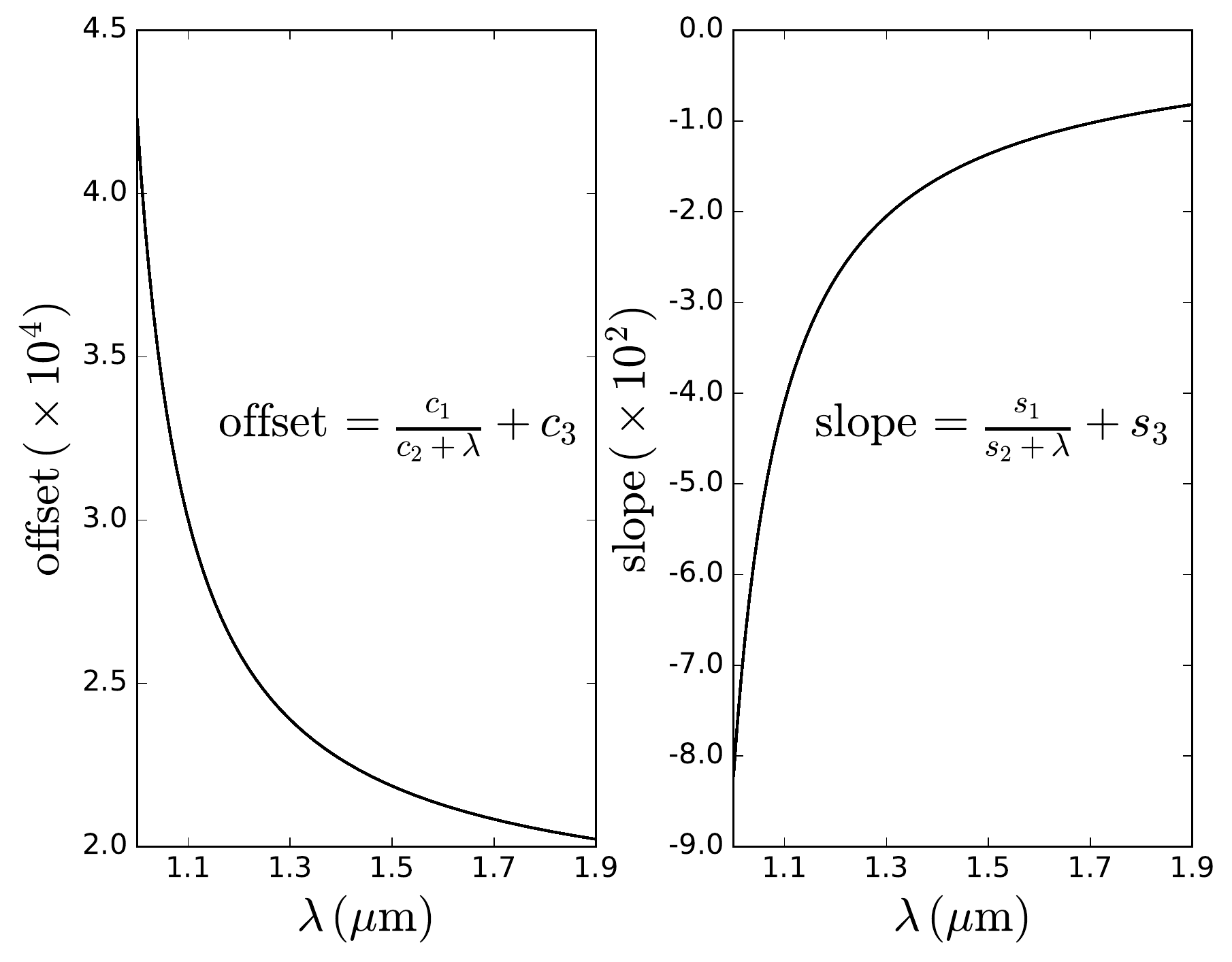}
	\caption{The offset and slope of the wavelength-dependent photon trajectories (w.d.p.t.) as functions of wavelength.\vspace{2mm}}
	\label{fig:offset-slope}
\end{figure}

For a given wavelength, equation \ref{wlcalib} represents a straight line across the detector, the line on which the photons of this particular wavelength move during the scan. In figure \ref{fig:offset-slope}, the offset and slope of these lines are plotted as functions of wavelength.

\subsubsection{Wavelength-dependent flat-field} \label{sub:flat-field}

We use the wavelength grid created in the previous section also to apply the wavelength-dependent flat-field, as described in the \textit{aXe User Manual version 2.3}. We can find the wavelength as a function of position by fitting a different two-dimensional function on the grid points:
\begin{equation}
	\lambda = \kappa_0 + \kappa_1 x_\lambda + \kappa_2 y_\lambda + \kappa_ 3 x_\lambda^2 + \kappa_4 x_\lambda y_\lambda + \kappa_5 y_\lambda^2
	\label{wlmap}
\end{equation}

The wavelength-dependent flat-field for each pixel ($x, y$), is then:
\begin{equation}
		F(x, y) = \sum_{i=0}^{i=3} F_i(x,y) \left( \frac{\lambda(x,y) - \lambda_\mathrm{min}}{\lambda_\mathrm{max} - \lambda_\mathrm{min}} \right) ^i
	\label{flat}
\end{equation}
\noindent where $F_i$ are the different extension arrays and $\lambda_\mathrm{min}, \lambda_\mathrm{max}$, the wavelength coefficients provided in the flat-field cube \textit{WFC3.IR.G141.flat.2.fits} \citep{flat}.
% Wavelength calibration

% Extraction of 1D spectra
\subsection{Extraction of 1D spectra} \label{sub:extraction}

\begin{figure}
	\centering
	\includegraphics[width=\columnwidth]{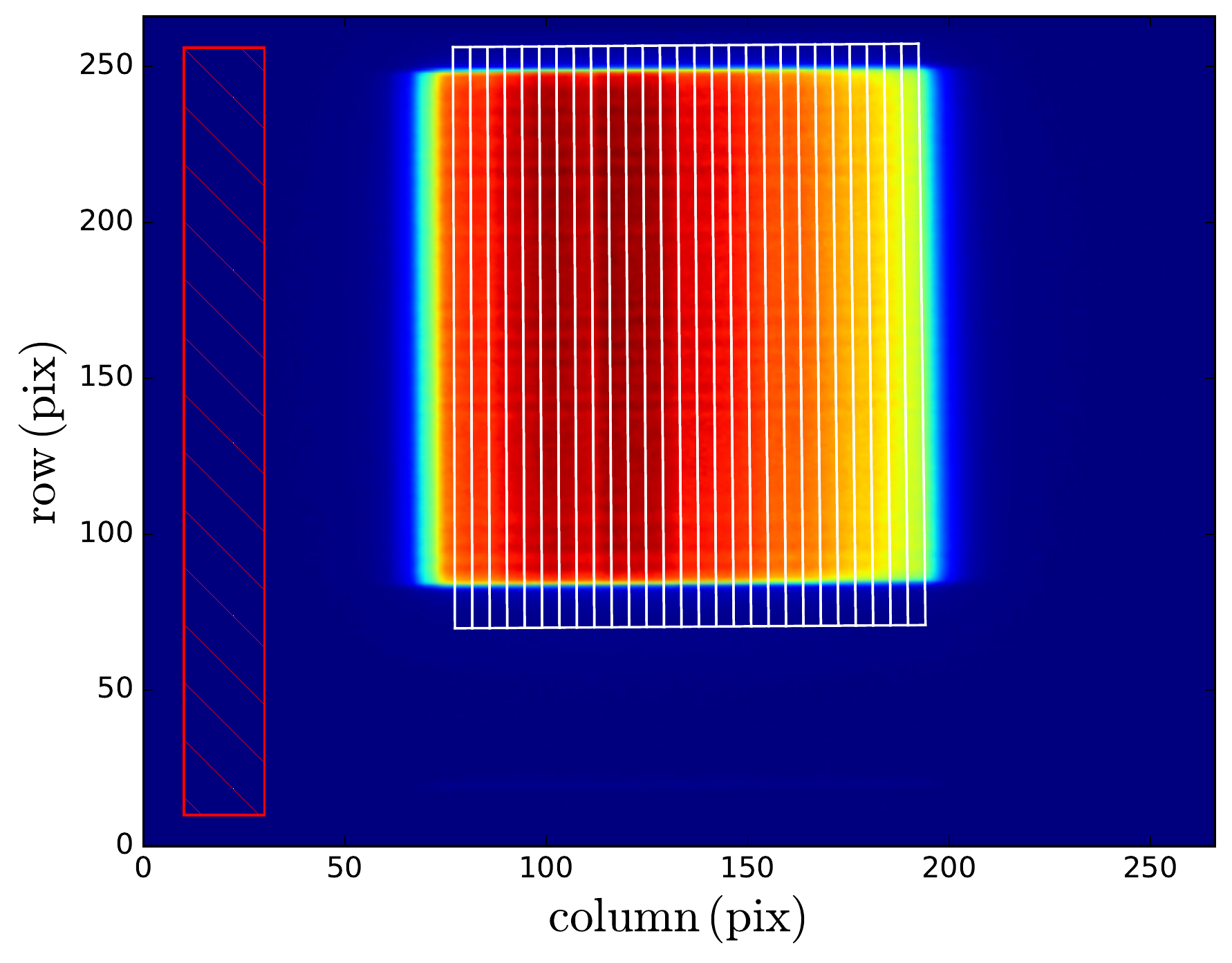}
	\caption{Photometric apertures for the different wavelength channels and the area (red square) from which the sky background ratio is estimated.}
	\label{fig:aperture}
\end{figure}

The 1D spectra are extracted from apertures of quadrangular shape, specifically calculated for each wavelength bin ($\lambda_1 - \lambda_2$) per frame (Figure \ref{fig:aperture}). The left and right edges of each quadrilateral are given by the \hyperlink{w.d.p.t.}{w.d.p.t.} (Equation \ref{wlcalib}) for $\lambda=\lambda_1$ and $\lambda=\lambda_2$, respectively. The upper and lower edges are given by the spectrum trace (Equation \ref{trace}) for $y^*=y_1+\Delta y_i$ and $y^*=y_1+\Delta y_i + l_i + y_2$ , respectively. $\Delta y_i$ is the vertical position shift and $l_i$ is the scan length of each spatially scanned spectrum, as calculated in Section \ref{sub:position}. The values for $y_1$ and $y_2$ are chosen in order to correspond to 15 pixels below and 7 pixels above the spatially scanned spectrum (for this dataset $y_1=442.5$ and $y_2=7$, pixels).

An issue concerning the extraction method, is that we have to take into account fractional pixels at the edges of the photometric apertures. As a first approximation we used the fraction of the pixel area inside the extraction aperture. While testing this method, we concluded that this approach intensifies the wavelength-dependent systematics that are caused be the horizontal shifts and the low spectral resolution of the spectrum (Section \ref{sub:spectral}). A better approach for those pixels is a second-order 2D polynomial distribution of the flux. The coefficients of this 2D function are calculated analytically so that its integral inside the pixel of interest and inside each surrounding pixel, are equal to their flux levels. We can then calculate the analytic integral of this function inside the common area of the pixel that we want to split and the extraction aperture.

\begin{figure}
	\centering
	\includegraphics[width=\columnwidth]{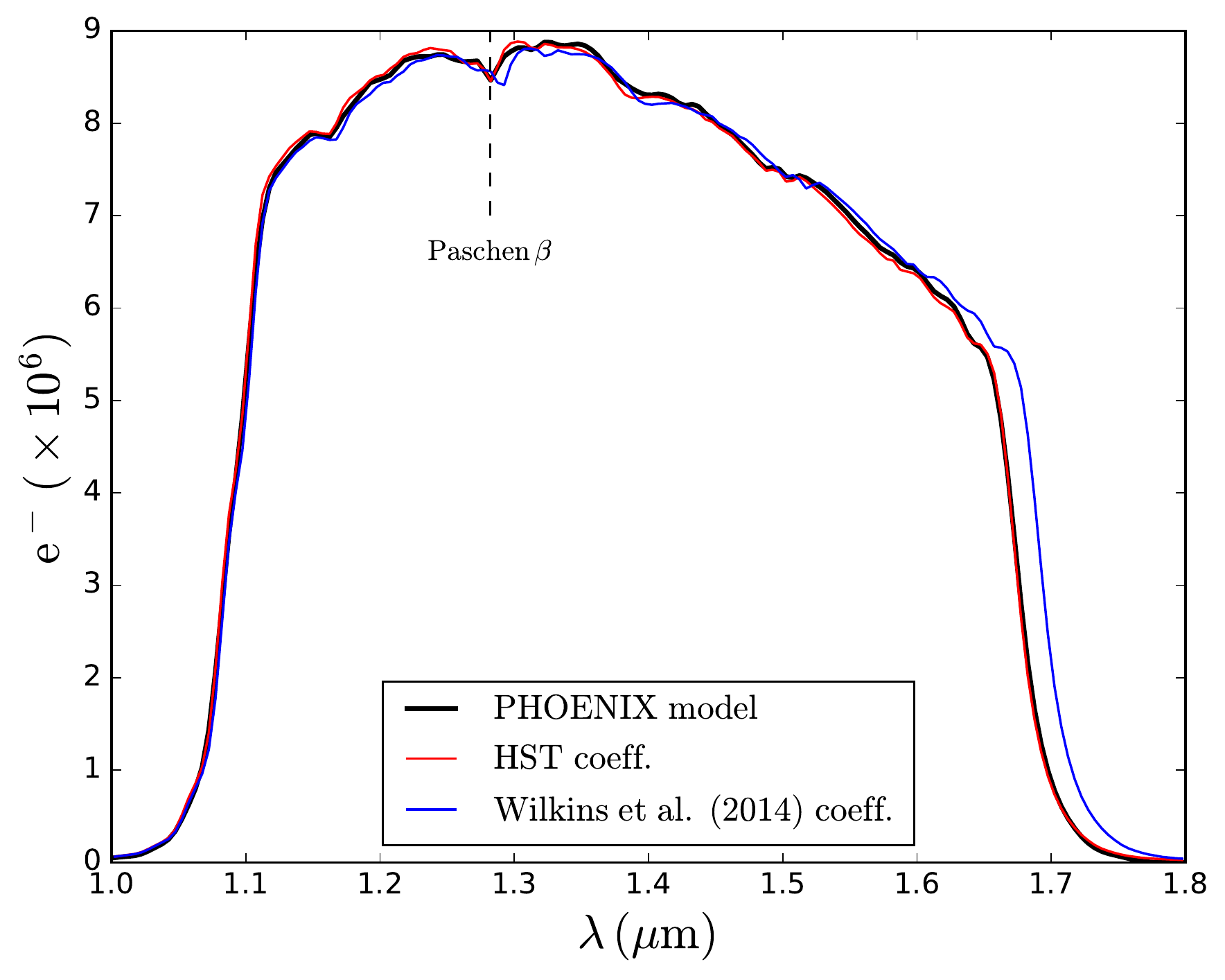}
	\caption{Extracted 1D spectrum using our method with the \hst\ calibration coefficients (red), and those proposed by \cite{Wilkins2014} (blue). For comparison, the PHOENIX model of the host star scaled by sensitivity curve of the G141 grism (black), and the position of the Paschen $\beta$ line.}
	\label{fig:sensitivity}
\end{figure}

\begin{figure}
	\centering
	\includegraphics[width=\columnwidth]{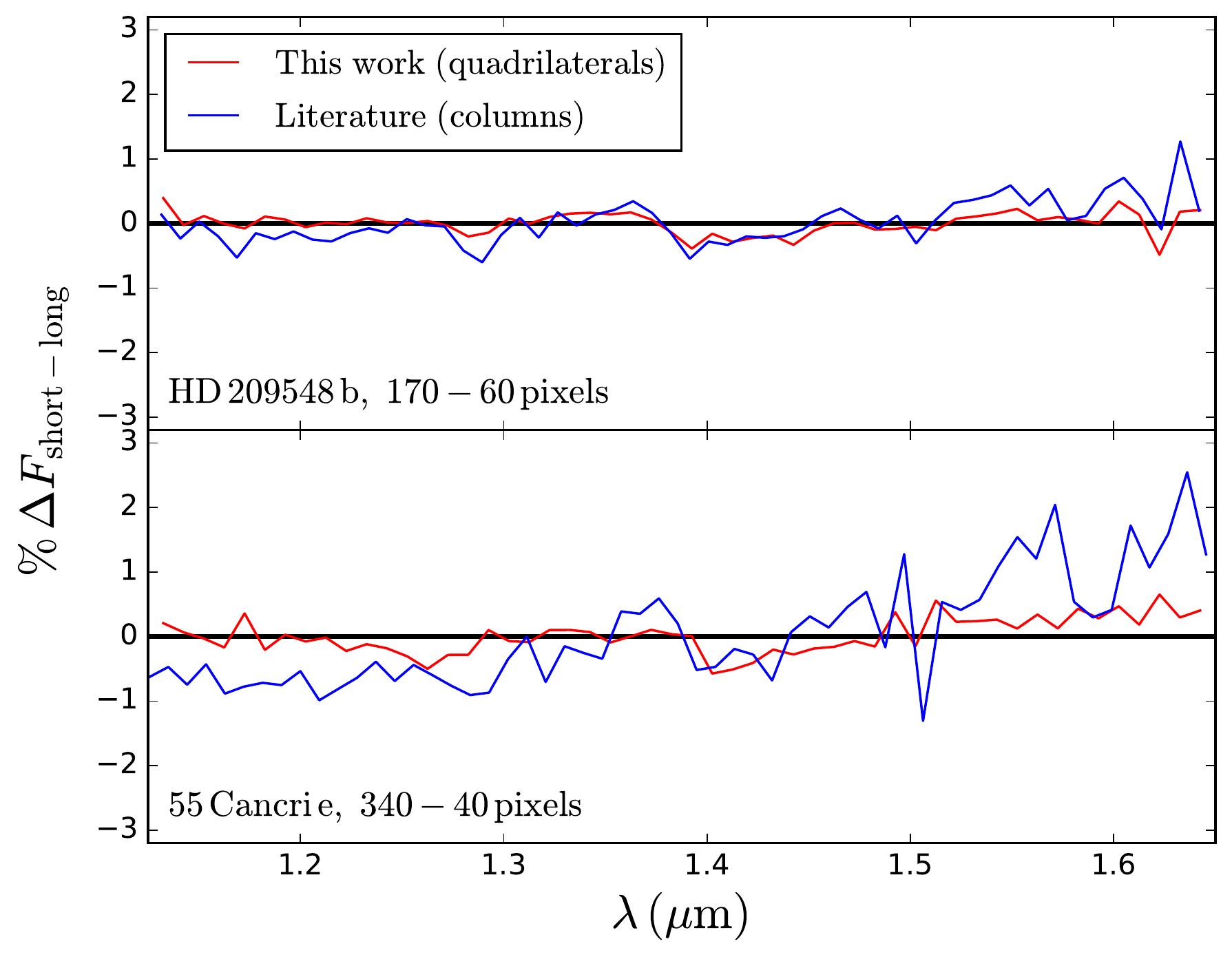}
	\caption{Percentage difference between the flux rate extracted from a short, intermediate, scan and the long, final, scan of the same exposure, using our method (red) and the sum along the columns (blue). At the top, the case of \planet, studied in this work, and at the bottom, the case of 55\,Cancri\,e \citep{Tsiaras2016}, a much longer scan. In both cases, our method gives three times better rms.}
	\label{fig:up-the-ramp}
\end{figure}

Overall, by taking into account simultaneously the geometrical distortions (dispersion variations across the scanning direction and inclined spectrum) and the positional shifts (horizontal and vertical), our calibration and extraction pipeline has two main advantages:
\begin{enumerate}
\item the photometric apertures are consistent with the geometric structure of the spatially scanned spectra across the detector, improving the consistency between short and long scans by three times, compared to summing along the columns (Figure \ref{fig:up-the-ramp}),
\item the extracted 1D spectra are consistent with the sensitivity curve of the G141 grism (Figure \ref{fig:sensitivity}), suggesting that there is no need to change the \hst\ calibration coefficients (used in equations \ref{trace} and \ref{wlsolution}), as proposed by \cite{Wilkins2014}.
\end{enumerate}

We will make the complete for reduction, calibration and extraction, available to the community in the near future. Meanwhile, all our intermediate results (reduced data and light-curves) are available for direct comparisons with other methods\footnote{\url{http://zuserver2.star.ucl.ac.uk/~atsiaras/wfc3/}}.

% Extraction of 1D spectra
% DATA ANALYSIS

% LIGHT-CURVE ANALYSIS
\section{LIGHT-CURVE ANALYSIS}

% Fitting the white light-curve
\subsection{Fitting the white light-curve} \label{sub:white_lc}

\begin{figure}
	\centering
	\includegraphics[width=\columnwidth]{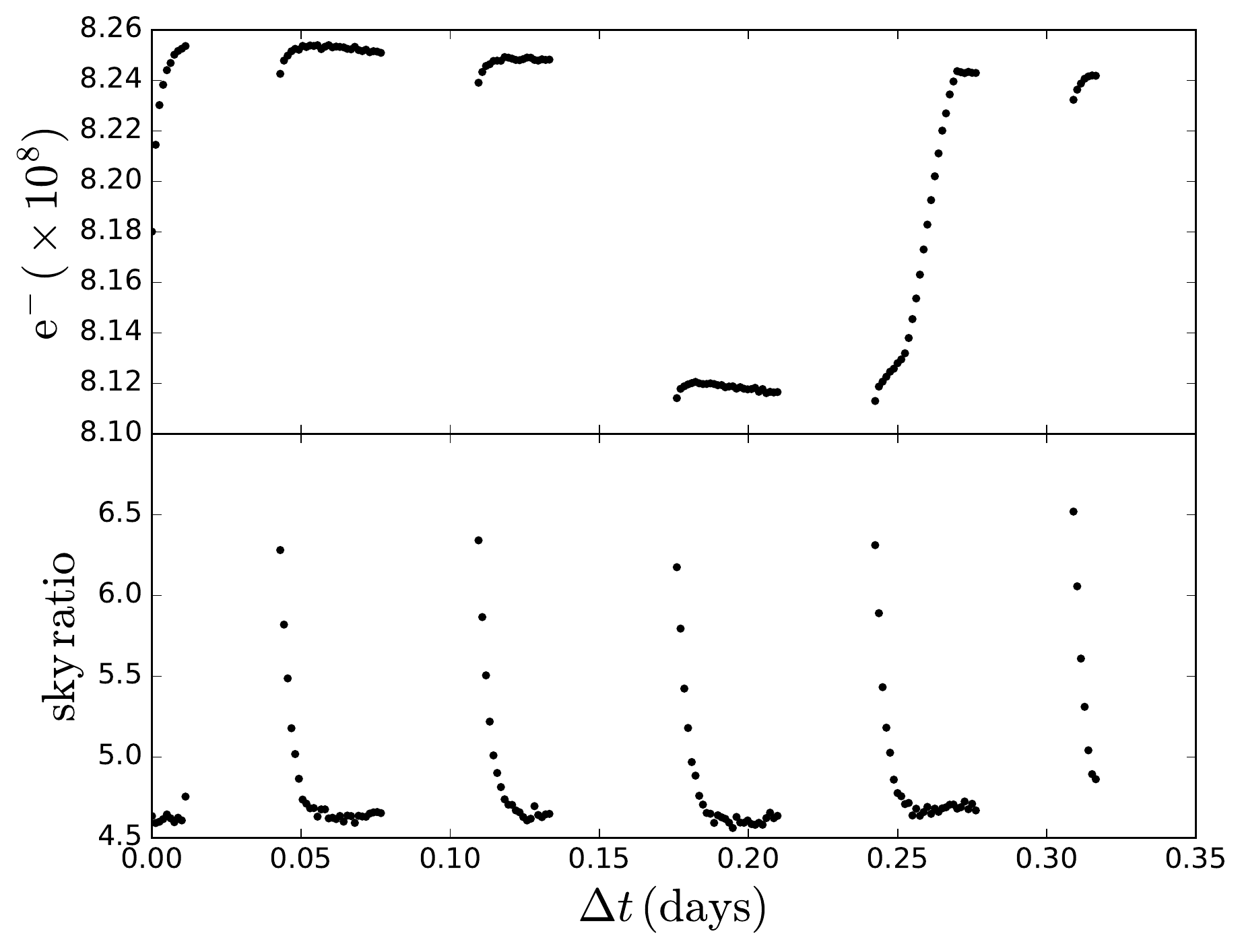}
	\caption{Raw white light-curve and sky background relatively to the master sky frame.}
	\label{fig:raw_white_lc}
\end{figure}

Having extracted the 1D spectra from all the frames we produce the white and spectral light-curves. It is known from previous studies of observations with WFC3 in staring-mode \citep{Berta2012, Swain2013, Wilkins2014} and scanning mode \citep{Deming2013, Kreidberg2014a, Knutson2014a} that the infrared detector introduces two time-dependent systematics to the light-curves of bright sources like \system: one long-term (throughout the visit) with an approximately linear behavior and one short-term (throughout each \hst\ orbit) with an approximately exponential behavior. These systematics are commonly referred to as the ``ramps'', and can be easily seen in the raw white light-curve (Figure \ref{fig:raw_white_lc}) but are also present in the light-curves of all wavelength channels. In this data set, the long-term ramp can be approximated by a linear function only after the third orbit. We do not include the first two orbits in our analysis as a wrong fitting of the behavior of the instrument would introduce uncertainties to the final values of the transit parameters. 

To correct these systematics we fit a transit model, $F(t)$, multiplied by a normalisation factor, $n_\mathrm{w}$, and an instrumental systematics function, $R(t)$, \citep{Kreidberg2014a, Kreidberg2014b, Stevenson2014, Kreidberg2015}:
\begin{equation}
	R(t) = (1 - r_a (t-T_0))(1-r_{b1} e^{-r_{b2} (t-t_\mathrm{o})})
	\label{eq:ramp_function}
\end{equation}

\noindent where $t$ is time, $T_0$ is the mid-transit time, $t_\mathrm{o}$ is the time when each orbit starts, $r_a$ is the slope of the linear, long-term ``ramp'' and ($r_{b1},r_{b2}$) are the coefficients of the exponential short-term ``ramp''.

For the transit part of the light-curve, we use our numerical model, which is written entirely in Python\footnote{\url{https://github.com/ucl-exoplanets/pylightcurve}}. It returns the relative flux, $F(t)$, as a function of the limb darkening coefficients, $a_n$, the $R_\mathrm{p}/R_*$ ratio and all the orbital parameters ($T_0, P, i, a/R_*, e, \omega$), based on the non-linear limb darkening model \citep{Claret2000} for the host star:
\begin{equation}
	I(a_n, r) = 1 - \sum_{n=1}^{n=4} a_n (1-(1-r^2)^{n/4})
	\label{eq:limb_darkening_law}
\end{equation}

We calculate the limb darkening coefficients by fitting an ATLAS model \citep{Kurucz1970, Howarth2011, Espinoza2015}. The ATLAS model is created using the stellar parameters in Table \ref{tab:parameters} and the sensitivity curve of the G141 grism, between 1.125 and 1.65\,$\mu$m (Table \ref{tab:fitting_white_lc}). We use a circular orbit and fix inclination and $a/R_*$ ratio to the values of Table \ref{tab:parameters}. Pre-selecting the values for the limb darkening coefficients and the orbital parameters is necessary, as the asymmetry in the light-curve (Figure \ref{fig:raw_white_lc}) does not allow us to constrain them from the data.

\begin{figure}
	\centering
	\includegraphics[width=\columnwidth]{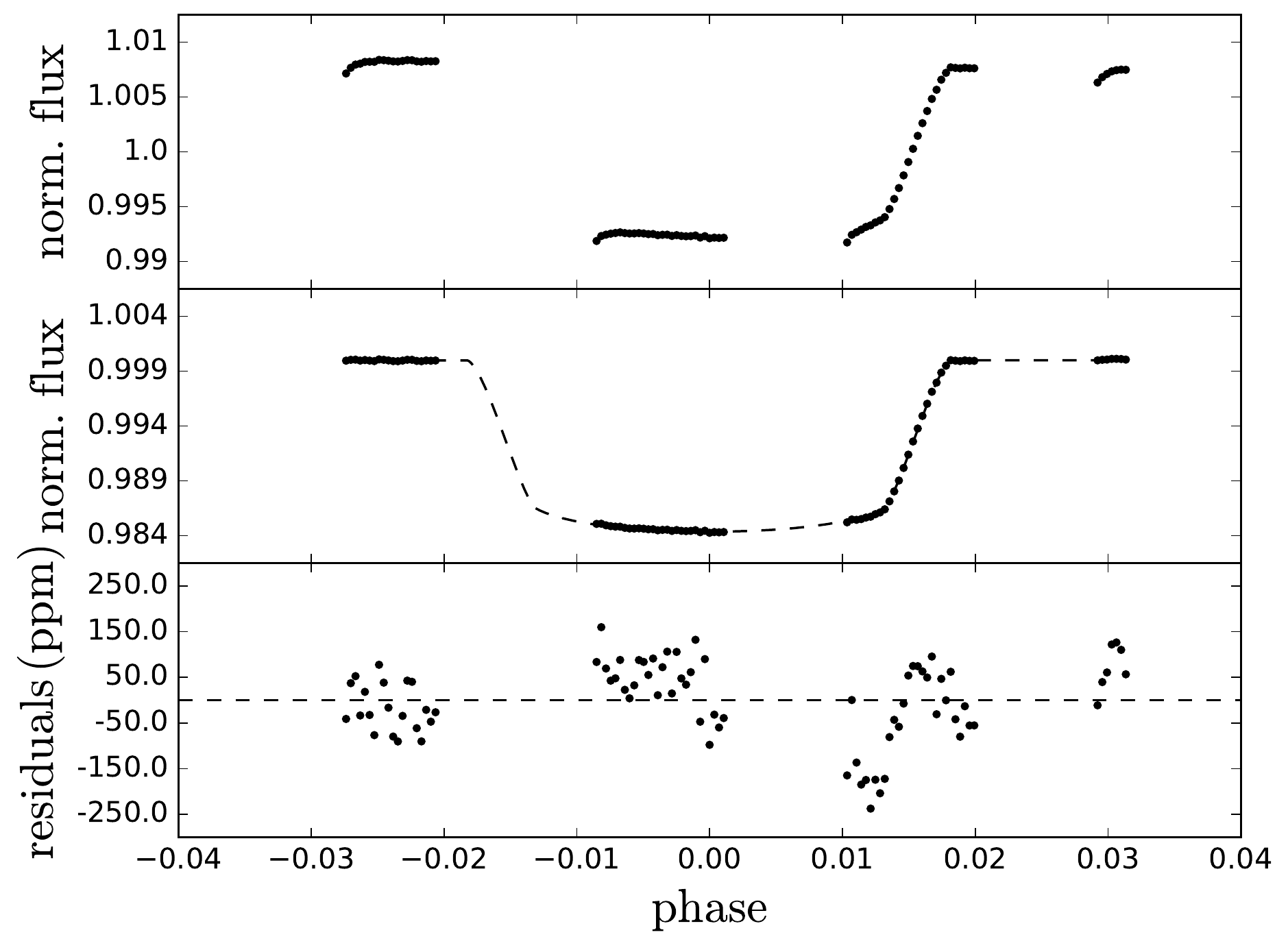}
	\caption{Top: Normalized raw white light-curve. Middle: white light-curve divided by the best-fit model for the systematics. Bottom: Fitting residuals, where we can see that the model fails to fit the egress. The most possible reasons for this behavior are either non-optimal orbital parameters, limb darkening coefficients or remaining systematics.}
	\label{fig:fitting_white_lc}
\end{figure}

As we can see at the bottom panel of figure \ref{fig:fitting_white_lc}, the residuals do not follow a Gaussian distribution at the transit egress. This behavior could be due to either non-optimal values used for the inclination and $a/R_*$ ratio or remaining systematics. For this reason we re-scale the uncertainties of the data points to the rms of the residuals and fit the light curve again. This increases the initial uncertainties approximately by three times. The fitting results and the final uncertainties can be found in Table \ref{tab:fitting_white_lc}. To verify the resulting spectrum and also reduce the uncertainties down to the noise floor and the residual error limit, a second measurement of the spectrum, time shifted to completing the phase coverage, would be required. 

\begin{table}
	\small
	\center
	\caption{White light-curve fitting results.}
	\label{tab:fitting_white_lc}
	\begin{tabular}{c | c}
		\hline \hline
		\multicolumn{2}{c}{Limb darkening coefficients (1.125 - 1.650 $\mu$m)}		\\ [0.1ex]
		\hline	
		$a_1$ 						& 0.608377						\\
		$a_2$						& $-$0.206186						\\
		$a_3$						& 0.262367						\\
		$a_4$						& $-$0.133129						\\ [1.0ex]
		
		\hline \hline
		\multicolumn{2}{c}{Fitted transit parameters}							\\
		\hline
		$T_0 \, \mathrm{(HJD)}$			& 2456196.28836\,$\pm$\,0.00005		\\
		$R_\mathrm{p}/R_*$				& 0.12079\,$\pm$\,0.00014				
	\end{tabular}
\end{table}

\begin{figure}
	\centering
	\includegraphics[width=\columnwidth]{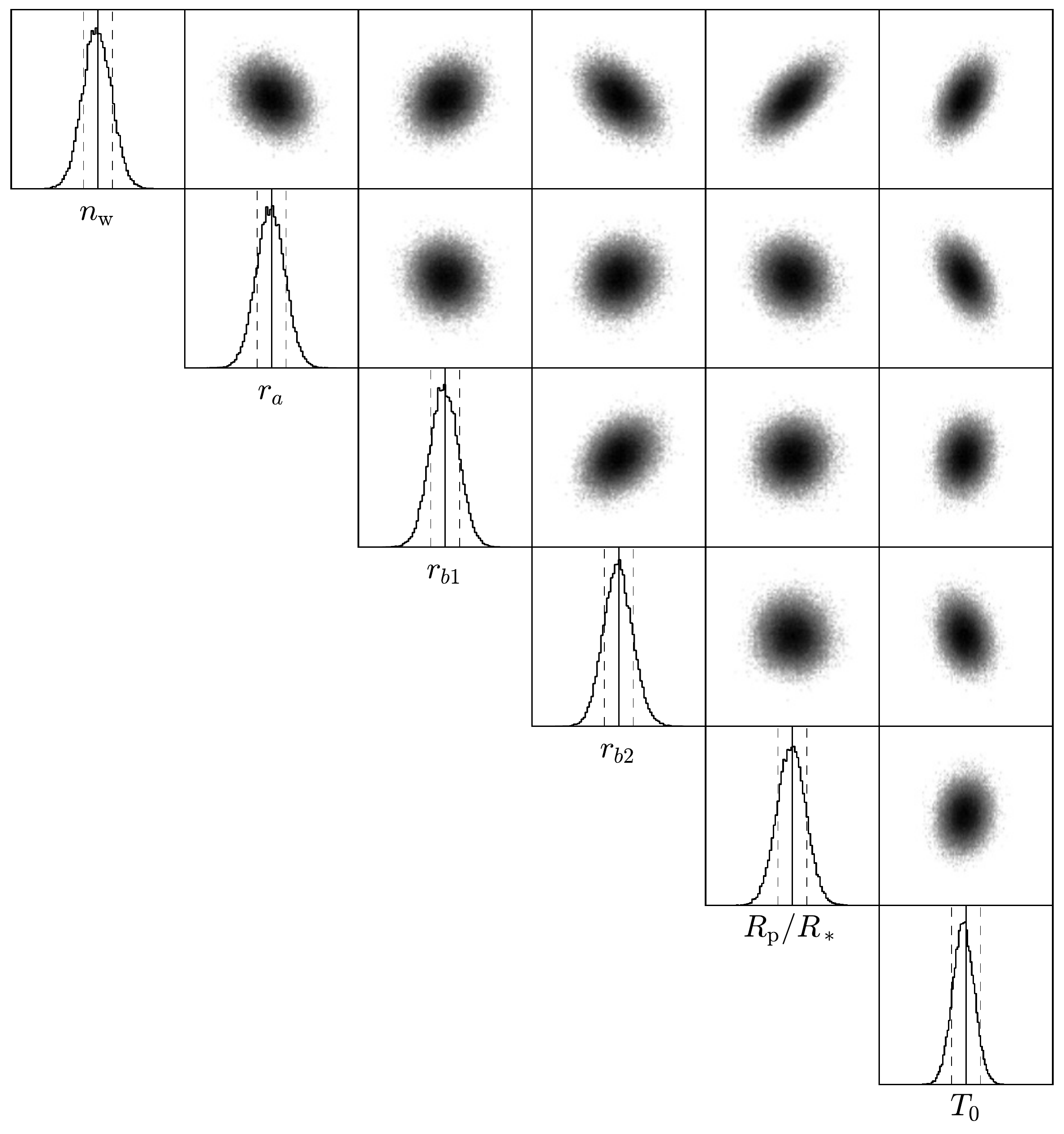}
	\caption{Correlations between the fitted systematics and transit parameters for the simultaneous fitting approach on all the data points. Apart from the expected correlation with the normalization factor, $R_\mathrm{p}/R_*$ ratio is not correlated with any of the three parameters that describe the systematics. On the contrary, $n_\mathrm{w}, r_a$ and $T_0$ are correlated with each other, due to the asymmetric distribution of the data points around $T_0$.}
	\label{fig:correlations_white_lc}
\end{figure}

The correlations between the fitted parameters are shown in Figure \ref{fig:correlations_white_lc}. We find no correlation between the $R_\mathrm{p}/R_*$ ratio and any of the ``ramp'' parameters, while $n_\mathrm{w}, r_a$ and $T_0$ are correlated with each other. These correlations are introduced by the asymmetry in the light-curve, as there is no constrain for the time of ingress. We do not find such correlations in the case of simulated symmetric light-curves.
% Fitting the white light-curve

% Fitting the spectral light-curves
\subsection{Fitting the spectral light-curves} \label{sub:spectral}

The wavelength bins are selected in such a way that: a) the total flux is equally distributed among the bins, to have an approximately uniform S/N and b) we avoid splitting the spectrum an wavelengths where the stellar spectrum has significant variations (1.165, 1.282, 1.372, and 1.502\,$\mu$m). We calculate the limb darkening coefficients for each spectral light-curve using the ATLAS model, the stellar parameters in Table \ref{tab:parameters} and the sensitivity curve of the G141 grism inside the boundaries of each wavelength bin. 

To extract the planetary spectrum from the spectral light-curves, we follow two approaches similar to those described in \cite{Kreidberg2014a}: a) we fit each spectral light-curve in the same way as the white one \textemdash \, i.e. fitting a wavelength-dependent normalisation factor, $n_\lambda$, a wavelength-dependent instrumental systematics function, $R(\lambda, t)$, and a wavelength-dependent transit model, $F(\lambda, t)$  \textemdash \, (method 1) and  b) we divide each spectral light-curve by the white one and then fit for a wavelength-dependent normalisation factor, $n_\lambda$, a wavelength-dependent linear slope, linear with time, $1 + \chi_\lambda (t-T_0)$, and a wavelength-dependent relative transit model, $F(\lambda, t)/F_\mathrm{w}(t)$, (method 2):
\begin{equation}
	\begin{split}
		\text{method 1: \, }	&	n_\lambda R(\lambda, t) F(\lambda, t) \\ 
		\text{method 2: \, }	&	n_\lambda (1 + \chi_\lambda (t-T_0)) (F(\lambda, t)/F_\mathrm{w}(t))
	\end{split}
	\label{eq:sectral_ramp}
\end{equation}

\noindent where $t$ is time, $T_0$ is the mid-transit time from Table \ref{tab:fitting_white_lc}, $\chi_\lambda$ is the coefficient of the wavelength-dependent linear slope, and $F_\mathrm{w}(t)$ is the best-fit model on the white light-curve (Section \ref{sub:white_lc}). In all the $F(\lambda, t)$ models, the only free parameter is the $R_\mathrm{p}/R_*$ ratio, while the other parameters are the same as in the white light-curve. Concerning the uncertainties, we re-scale the uncertainties of the data points to the rms of the residuals and fit again, in the same way as for the white light-curve.
% Fitting the spectral light-curves
% LIGHT-CURVE ANALYSIS

% ATMOSPERIC RETRIEVAL
\section{ATMOSPERIC RETRIEVAL} \label{sec:retrieval}

We used the nested sampling algorithm implemented in \taurex\ \citep{Waldmann2015b, Waldmann2015a} to fully explore the parameter space and find the best fit to the WFC3 spectrum. Because of the limited number of data points in the observed spectrum, in order to significantly reduce the parameter space we parametrise the atmosphere assuming an isothermal profile, with constant molecular abundances as a function of altitude. The fitted parameters are the temperature, the molecular abundances for the different species, the mean molecular weight, the radius at 10 bar, and the cloud top pressure \textemdash \ i.e. the pressure at which the cloud starts to be opaque. The cloud model used assumes an opaque and uniformly distributed cloud deck defined at a given pressure beyond which electromagnetic radiation is blocked at all wavelengths. We consider a broad range of absorbing molecules, including H$_2$O, HCN, NH$_3$, CH$_4$, CO$_2$, CO, NO, SiO, TiO, VO, H$_2$S, C$_2$H$_2$. 

We fit for the individual molecular abundances, assuming the bulk composition of the atmosphere to be made by a mixture of 85\% hydrogen and 15\%  helium. We then couple the mean molecular weight to the atmospheric composition.  We consider uniform priors for the molecular volume mixing ratios ranging between $10^{-12}$ and $10^{-2}$. This prior is justified by the fact that in hot Jupiters the absolute abundances of absorbing gases are significantly smaller compared to the H$_2$O and He content. We also assume uniform priors for the temperature ($T = 1000 - 1800 \,K$), 10 bar radius ($R = 1.3 - 1.4 \, R_\mathrm{Jup}$) and cloud top pressure ($P_\mathrm{cloud} = 10^{-5} - 10^{-1} \, \mathrm{Pa}$). We run two retrievals, the first one including 12 molecules and aimed at identifying the most likely trace gases, and the second one including only the molecules identified in the first run, aimed at fully mapping the parameter space and at investigating the degeneracy of the model. 
% ATMOSPERIC RETRIEVAL

%RESULTS
\section{RESULTS} \label{sec:results}

\begin{figure}
	\centering
	\includegraphics[width=\columnwidth]{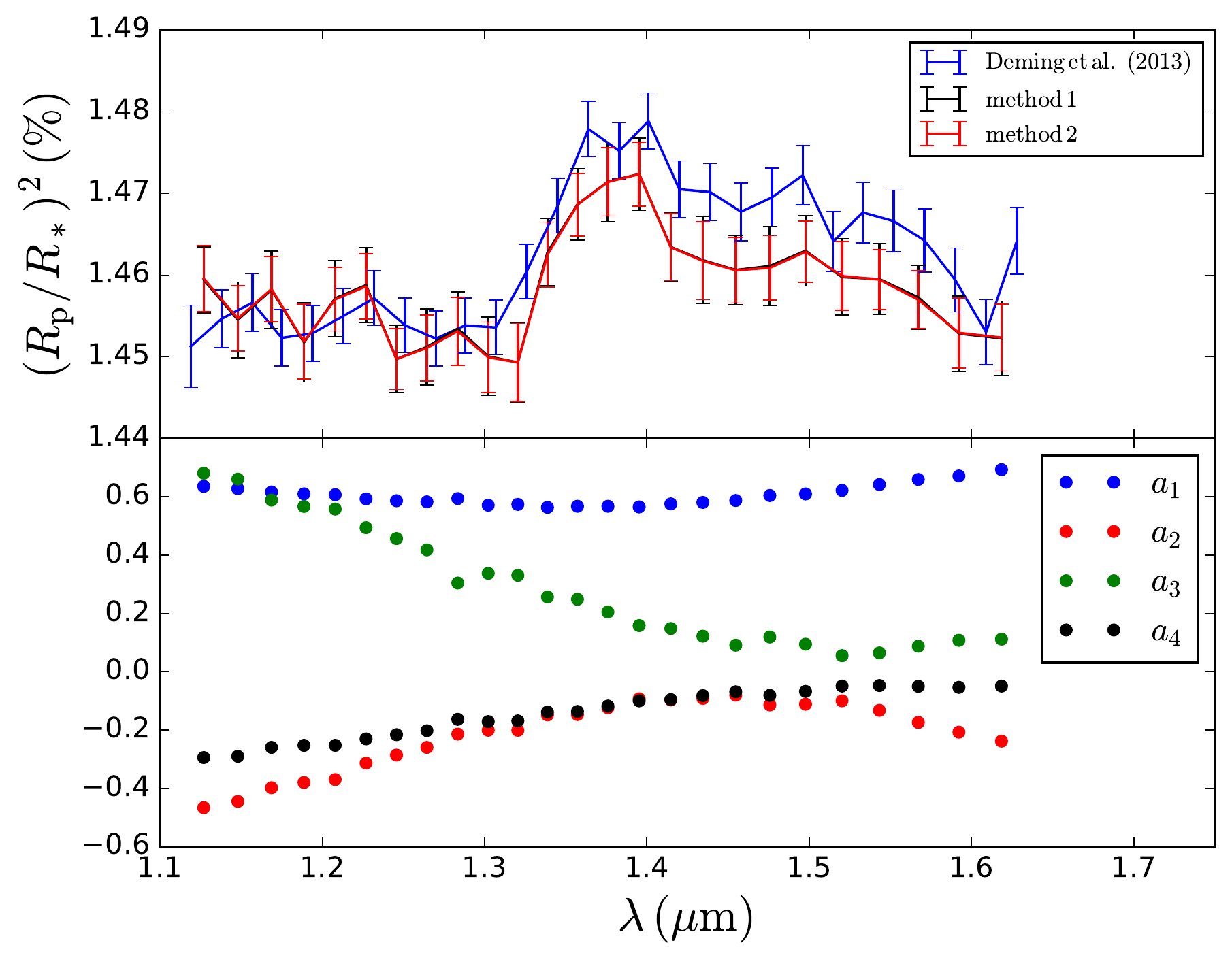}
	\caption{Top: Comparison between the extracted spectra using the two different methods, and the spectrum from \cite{Deming2013}. Bottom: Limb darkening coefficients ($a_{1-4}$) as functions of wavelength.}
	\label{fig:spectrum}
\end{figure}

The limb darkening coefficients, $a_{1-4}$, fitted on the ATLAS model, and the final measurements of the transit depth, $(R_\mathrm{p}/R_*)^2$, as a function of wavelength, $\lambda$, are presented in Figure \ref{fig:spectrum}. The results from the two methods agree within 3\,ppm, while the uncertainties are of the level of 40\,ppm. However, the uncertainties in method 2 are improved by 10\% compared to method 1. Method 2 performs better because the ``ramp'' model, used in method 1, cannot reproduce perfectly the real systematics. We therefore use the results from method 2 (Table \ref{tab:spectrum}) in the spectral retrieval.

We found the slope of the long-term ``ramp'' to be wavelength-dependent, as well as, the term $\chi_\lambda$ in model 2. This behaviour supports the hypothesis that while the exponential ``ramp'' is a common-mode between the white and the spectral-light curve, the linear ``ramp'' is not, as seen in previous \wfc\ observations \citep[e.g.][]{Deming2013, Fraine2014, Wilkins2014, Kreidberg2015}.

Figure \ref{fig:spectrum} plots, also, the spectrum obtained by \cite{Deming2013}. While the two spectra include the same features, we can see that at longer wavelengths there is a systematic difference. This difference could be caused by the geometric distortions, that are stronger at longer wavelengths, or buy the different way of taking into account the limb darkening coefficients.

\begin{table*}
	\small
	\center
	\caption{Limb darkening coefficients $a_{1-4}$ and transit depth for the different wavelength channels.}
	\label{tab:spectrum}
	\begin{tabular}{r l | c c c c | c}
		\hline \hline
		\multicolumn{2}{c |}{$\lambda_1- \lambda_2 \, (\mu \mathrm{m})$} 	 & $a_1$ 	 & $a_2$ 	 & $a_3$	 & $a_4$ 	 & $(R_\mathrm{p}/R_*)^2 \, \mathrm{(ppm)}$ 		 \\ [0.1ex]
		\hline
		1.1165 	 & 1.1375 					 & 0.635743 	 & -0.466435 	 & 0.680708 	 & -0.294517 	 & 14596 $\pm$ 41 	 \\
		1.1375 	 & 1.1585 					 & 0.627663 	 & -0.444932 	 & 0.660394 	 & -0.290206 	 & 14547 $\pm$ 40 	 \\
		1.1585 	 & 1.1790 					 & 0.616173 	 & -0.397988 	 & 0.58849 	 & -0.25968 	 & 14583 $\pm$ 40 	 \\
		1.1790 	 & 1.1985 					 & 0.609875 	 & -0.379926 	 & 0.566774 	 & -0.252676 	 & 14518 $\pm$ 46 	 \\
		1.1985 	 & 1.2175 					 & 0.606905 	 & -0.369955 	 & 0.557771 	 & -0.252625 	 & 14570 $\pm$ 39 	 \\
		1.2175 	 & 1.2365 					 & 0.592654 	 & -0.313497 	 & 0.494043 	 & -0.230553 	 & 14586 $\pm$ 40 	 \\
		1.2365 	 & 1.2550 					 & 0.586298 	 & -0.286056 	 & 0.456591 	 & -0.216029 	 & 14497 $\pm$ 37 	 \\
		1.2550 	 & 1.2740 					 & 0.582601 	 & -0.259779 	 & 0.417679 	 & -0.202415 	 & 14511 $\pm$ 41 	 \\
		1.2740 	 & 1.2930 					 & 0.593752 	 & -0.214099 	 & 0.304066 	 & -0.163396 	 & 14531 $\pm$ 42 	 \\
		1.2930 	 & 1.3115 					 & 0.570891 	 & -0.201033 	 & 0.337262 	 & -0.17122 	 & 14499 $\pm$ 43 	 \\
		1.3115 	 & 1.3295 					 & 0.573679 	 & -0.201341 	 & 0.330334 	 & -0.168801 	 & 14493 $\pm$ 48 	 \\
		1.3295 	 & 1.3480 					 & 0.563602 	 & -0.148149 	 & 0.256247 	 & -0.138194 	 & 14625 $\pm$ 40 	 \\
		1.3480 	 & 1.3665 					 & 0.567258 	 & -0.146917 	 & 0.248296 	 & -0.136497 	 & 14686 $\pm$ 38 	 \\
		1.3665 	 & 1.3855 					 & 0.567304 	 & -0.123792 	 & 0.204782 	 & -0.117829 	 & 14714 $\pm$ 42 	 \\
		1.3855 	 & 1.4050 					 & 0.565076 	 & -0.0927791 	 & 0.158153 	 & -0.0998947 	 & 14724 $\pm$ 39 	 \\
		1.4050 	 & 1.4245 					 & 0.575439 	 & -0.0971152 	 & 0.148165 	 & -0.0955357 	 & 14634 $\pm$ 41 	 \\
		1.4245 	 & 1.4445 					 & 0.580506 	 & -0.0916465 	 & 0.121781 	 & -0.0819971 	 & 14618 $\pm$ 48 	 \\
		1.4445 	 & 1.4650 					 & 0.587064 	 & -0.0803535 	 & 0.0907723 	 & -0.0687593 	 & 14606 $\pm$ 40 	 \\
		1.4650 	 & 1.4865 					 & 0.604201 	 & -0.113934 	 & 0.119031 	 & -0.0810911 	 & 14609 $\pm$ 39 	 \\
		1.4865 	 & 1.5090 					 & 0.609562 	 & -0.111526 	 & 0.0943074 	 & -0.0675727 	 & 14629 $\pm$ 37 	 \\
		1.5090 	 & 1.5315 					 & 0.62174 	 & -0.0998157 	 & 0.0551344 	 & -0.0490792 	 & 14599 $\pm$ 42 	 \\
		1.5315 	 & 1.5550 					 & 0.641863 	 & -0.132576 	 & 0.0645796 	 & -0.0474349 	 & 14595 $\pm$ 37 	 \\
		1.5550 	 & 1.5795 					 & 0.659312 	 & -0.174048 	 & 0.0873337 	 & -0.0501129 	 & 14570 $\pm$ 35 	 \\
		1.5795 	 & 1.6050 					 & 0.671289 	 & -0.207513 	 & 0.107652 	 & -0.0535945 	 & 14529 $\pm$ 43 	 \\
		1.6050 	 & 1.6320 					 & 0.692984 	 & -0.238165 	 & 0.111607 	 & -0.0491783 	 & 14524 $\pm$ 41 	 \\
	\end{tabular}
\end{table*}

\begin{figure}
	\centering
	\includegraphics[width=\columnwidth]{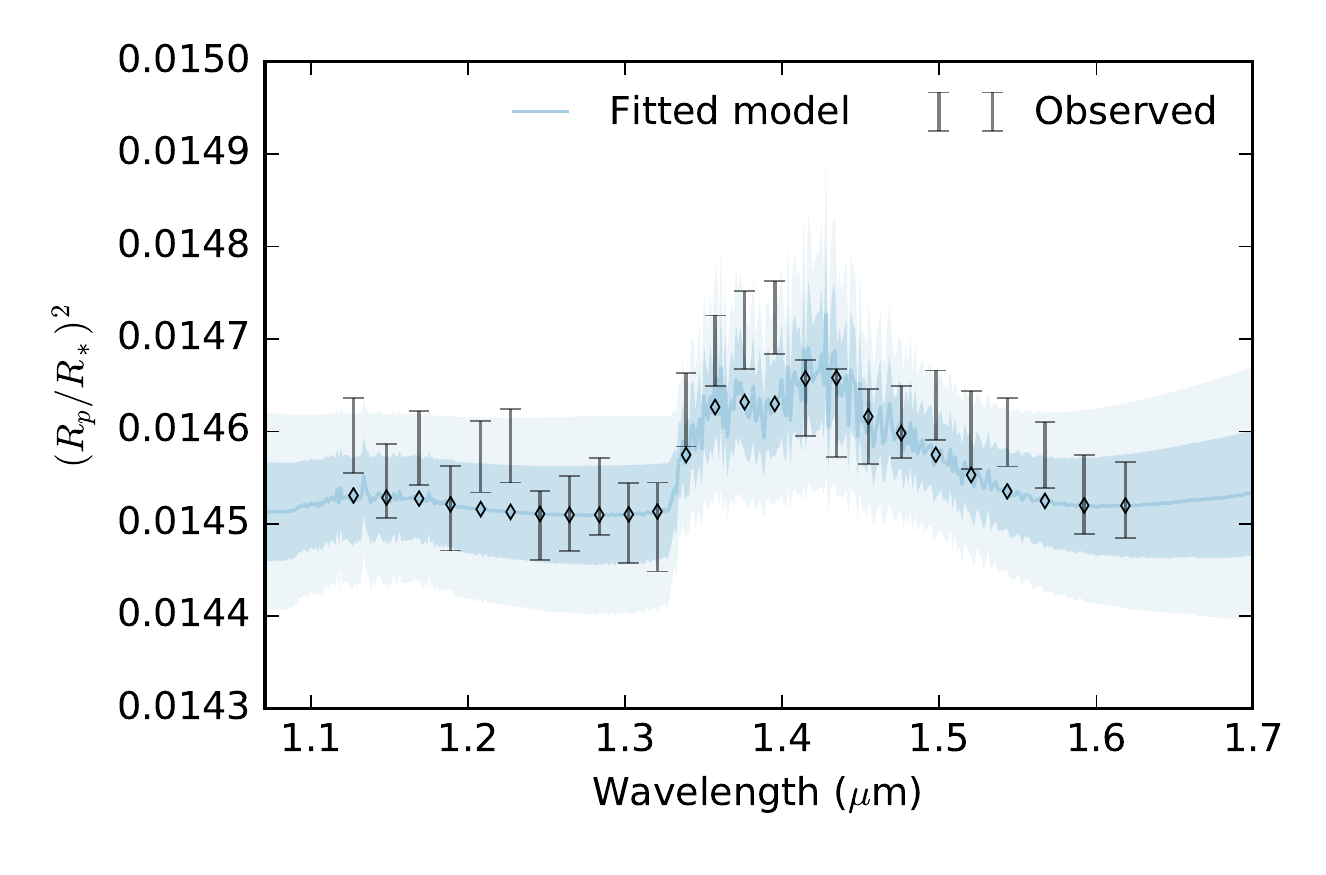}
	\caption{Infrared transmission spectrum of HD209458 b (black error bars), best fit obtained with the second retrieval containing H$_2$O and clouds (blue line).  The shaded regions show the the 1 and 2$\sigma$ confidence intervals in the the retrieved spectrum. }
	\label{retrievalfigure}
\end{figure}

\begin{figure}
	\centering
	\includegraphics[width=\columnwidth]{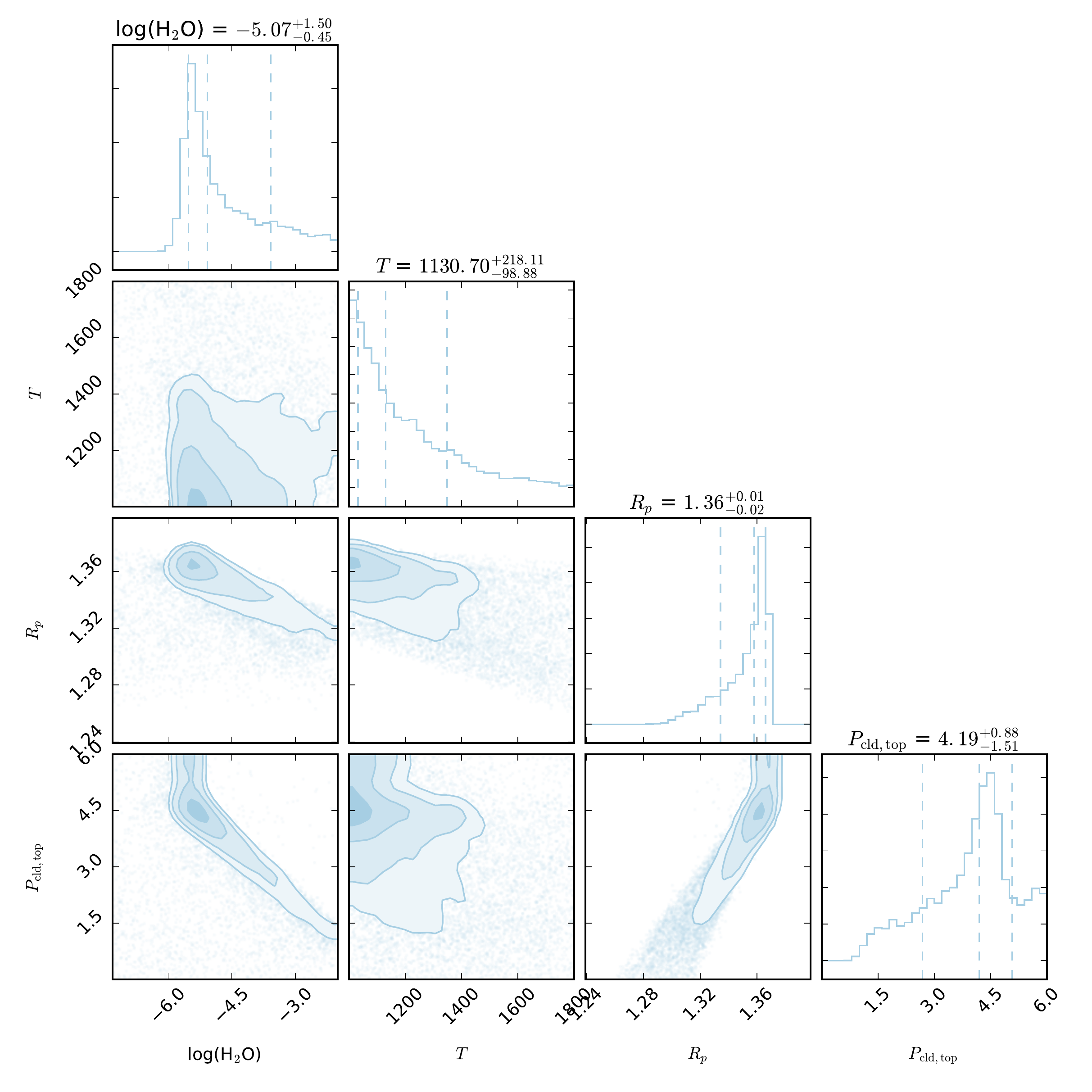}
	\caption{Posterior distributions of the second spectrum retrieval including H$_2$O and clouds.}
	\label{posterior}
\end{figure}

Figures \ref{retrievalfigure} and \ref{posterior} show the best fits to the spectrum obtained with \taurex\ and the posterior distributions of the second spectral retrieval, respectively. The first retrieval including all molecules shows that water is the strongest and most likely absorber, explaining the broad absorption feature at $\approx 1.35\,\mu$m. No other molecules seem to contribute to the overall spectrum, while clouds may be present to explain the flat spectrum seen between 1.1 and 1.3 $\mu$m. These results are in agreement with the previous analysis of this data set \citep{Deming2013}.

We therefore run the second retrieval including only H$_2$O and clouds. Figure \ref{retrievalfigure}  shows the best fit to the data corresponding to the maximum a posteriori solution of this Bayesian retrieval, while Figure \ref{posterior} shows the posterior distributions of this retrieval. We find that the retrieved absolute abundances of H$_2$O is $3\times10^{-6}-3\times10^{-4}$. However, the posterior distributions (Figure \ref{posterior}) shows that this parameter is highly degenerate with the cloud top pressure and the the 10 bar radius. It is therefore impossible with these data alone to constrain the absolute abundances of this absorber.

We found the 10 bar radius to be $1.36^{+0.01}_{-0.02}$\,$R_\mathrm{Jup}$. The posterior distributions also show that the data can be best explained by a cloud deck at 0.15 bar, but we note that the distribution is very broad (and degenerate with the other fitted parameters), and a solution without clouds or with lower-pressure clouds is also acceptable.

The first retrieval including all molecules has a global evidence $\log E = 209$, while the second retrieval including H$_2$O only has $\log E = 210$. Despite the global evidence of the H$_2$O-only retrieval being only marginally higher than that of the more complete model, this result shows that there is no statistical evidence that favor the presence of additional molecules in the spectrum.

Lastly, we find the atmospheric mean temperature peaking towards the lower edge of the prior. As the prior bounds are justified by the equilibrium temperature of the planet and a reasonable range of possible albedos, this result shows that the current model likely biases the retrieved temperature towards lower values.

%RESULTS

%DISCUSSION AND CONCLUSIONS
\section{DISCUSSION AND CONCLUSIONS} \label{sec:discussion}

The spatial scanning technique has improved the efficiency of the \wfc\ camera compared to the staring-mode observations, as it allows longer exposure times for bright targets, minimising the risk of saturating the sensitive detector. However, unlikely staring spectra, spatially scanned spectra are affected by the field-dependent characteristics of the \grism\ grism (dispersion variations across the scanning direction and inclined spectrum). In addition, scanning-mode observations include positional shifts (horizontal and vertical) that are an order of magnitude stronger than staring-mode observations.

We developed a new pipeline designed to minimise these effects on the spatially scanned spectra, by including alternative calibration and extraction techniques and using a coordinate system along the wavelength/scanning axes instead of the x/y axes of the detector. We found discrepancies up to 1\% the flux of the star between scans that differ by 100\,pixels in length. Consequently, for scan lengths of this range and beyond, the geometric distortions should be taken into account. We, also, found that the effect becomes stronger as the scan length increases. Our approach ensures the more efficient analysis of scans longer than 100\,pixels, and therefore of longer exposure times for bright targets, as demonstrated in \cite{Tsiaras2016}.

As a test case, we reanalysed the spatially scanned spectra during the transit of \planet. Because of the incomplete phase coverage, we were not able to investigate in more detail the effect of the time-dependent systematics on this data set. To further calibrate and verify the repeatability of the results obtained, a second, time-shifted, observation would be necessary.

The interpretation of the final spectrum with our retrieval code \taurex, confirms the presence of water vapour and also suggests the presence of clouds, in agreement with the literature. However, we note that it is not possible to determine the absolute abundances of this gas. To address this issue, additional infrared spectroscopic observations, over a broader wavelength range, are needed.
%DISCUSSION AND CONCLUSIONS

%AKNOLEDGEMENTS
\begin{acknowledgements}

This work was supported by STFC (ST/K502406/1) and the ERC projects ExoLights (617119) and ExoMol (267219)

\end{acknowledgements}
%AKNOLEDGEMENTS

%Appendix: TARGET POSITION
\appendix
\section{TARGET POSITION} \label{app:target}

During the calibration process, as explained in section \ref{sub:wavelength}, we calculate the physical position of the star on the detector using Equation \ref{pos_first}:
\begin{equation*}
	x^*=	x_0 + (507-0.5L) + \Delta x_\mathrm{off} + \Delta x_\mathrm{ref}
\end{equation*}

Here we give the values of all the parameters used in this calculation for all different sub-arrays, filters and apertures, apart from $x_0$ which is the result of fitting a 2D gaussian function to the direct image. \vspace{-5mm}

\begin{table*}[h!]
	
	\begin{minipage}[t]{0.48\textwidth}
	\small
	\centering
	
	\caption{Lengths of the different sub-arrays.
	\label{tab:sub-array_length}}
	\begin{tabular}[t]{l c | l c}
		\hline \hline
		array		& $L$ [pix]	& array		& $L$ [pix]	\\ [0.1ex]
		\hline 
		FULL	& 1024		& SUB128		& 128		\\ [-0.1ex]
		SUB512	& 512		& SUB64		& 64			\\ [-0.1ex]
		SUB256	& 256		&
	\end{tabular}

	\caption{Offsets from F140W filter.
	\label{tab:filter_offsets}}
	\begin{tabular}{l c | l c}
		\hline \hline
		filter		& $x_\mathrm{off}$ [pix]			& filter	& $x_\mathrm{off}$ [pix]			\\ [0.1ex]
		\hline 
		F098W	& 0.150				& F132N	& 0.039				\\ [-0.1ex]
		F140W	& 0.083				& F126N	& 0.264				\\ [-0.1ex]
		F153M	& 0.146				& F167N	& 0.196				\\ [-0.1ex]
		F139M	& 0.110				& F164N	& 0.169				\\ [-0.1ex]
		F127M	& 0.131				& F160W	& 0.136				\\ [-0.1ex]
		F128N	& 0.026				& F125W	& 0.064				\\ [-0.1ex]
		F130N	& 0.033				& F110W	& -0.073
	\end{tabular}
	\end{minipage}
	\begin{minipage}[t]{0.48\textwidth}
	\small
	\centering
	
	\caption{Chip reference pixels and pixel scales for the different WFC3 apertures (IR Channel). 
	\label{tab:wfc3_apertures}}
	\begin{tabular}{l c c}
		\hline \hline
		WFC3 apperture			& $x_\mathrm{ref}$ [pix]	& $x_\mathrm{scale}$ [''/pix]		\\ [0.1ex]
		\hline 
		IR						& 562.0			& 0.135601			\\ [-0.1ex]
		IR - G102 \& G141			& 497.0			& 0.135603			\\ [-0.1ex]
		IRSUB64/128/256/512		& 522.0			& 0.135470			\\ [-0.1ex]
		GRISM1024 - G102 \& G141	& 497.0			& 0.135603			\\ [-0.1ex]
		GRISM1024				& 497.0			& 0.135603			\\ [-0.1ex]
		GRISM512 - G102 \& G141	& 505.0			& 0.135504			\\ [-0.1ex]
		GRISM512 				& 505.0			& 0.135504			\\ [-0.1ex]
		GRISM256 - G102 \& G141	& 410.0			& 0.135508			\\ [-0.1ex]
		GRISM256				& 410.0			& 0.135508			\\ [-0.1ex]
		GRISM128/64 - G102		& 376.0			& 0.135476			\\ [-0.1ex]
		GRISM128/64 - G141		& 410.0			& 0.135474			\\ [-0.1ex]
		GRISM128/64				& 496.0			& 0.135404			
	\end{tabular}
	\end{minipage}

\end{table*}

$L$ is the size of the direct image (Table \ref{tab:sub-array_length}) and the correction $507-0.5L$ is needed to transform the calculated x-position from the coordinate system of the sub-array used for the direct image to that of the full detector array. It is a result of the fact that all the sub-arrays have the same centre as the full detector array. We also have to mention that the calibration coefficients do not take into account the reference pixels and so the centre is 507 instead of 512.

$\Delta x_\mathrm{off}$ is the difference in the centroid offsets along the x-axis between the filter used for the direct image and the filter F140W, as calculated by \cite{offsets} (Table \ref{tab:filter_offsets}). This correction is needed because all the calibration coefficients have been calculated relatively to direct images with the F140W filter.

Finally, $\Delta x_\mathrm{ref}$ is the difference in the chip reference pixels between the WFC3 aperture used for the direct image and the the aperture used for the dispersed image (Table \ref{tab:wfc3_apertures}). The reference pixel is the pixel where the given target coordinates are fixed by the telescope. It is usual to have a shift between the other filters and the two grisms, in oder for the spectrum to fit inside the sub-array. This correction also includes any shifts indicated by the observer through the POSTARG1 keyword in the fits file header (converted to pixels). Table \ref{tab:wfc3_apertures} contains most of the available apertures, a complete list can be found on the STScI website\footnote{\url{http://www.stsci.edu/hst/observatory/apertures/wfc3.html}}.
%Appendix: TARGET POSITION

%Appendix: WAVELENGTH GRID EQUATIONS
\section{WAVELENGTH GRID EQUATIONS} \label{app:grid}

In section \ref{sub:wavelength} we use equations \ref{wlpos} to calculate the position of the incoming photons $(x_\lambda, y_\lambda)$ as function of wavelength ($\lambda$) for a given physical position of the star on the full detector array $(x^*,y^*)$:
\begin{equation*}
	\begin{split}
		x_{\lambda} = 	&	x^* - \frac{b_\mathrm{t} a_\mathrm{t}}{a_\mathrm{t}^2 +1} + \frac{\lambda-b_\mathrm{w}}{a_\mathrm{w}} \cos [ \tan^{-1} (a_\mathrm{t}) ] \\
		y_{\lambda} =	&	a_\mathrm{t}(x_{\lambda} - x^*) + b_\mathrm{t} + y^*
	\end{split}
\end{equation*}

\noindent where $(a_{\mathrm{t}n},b_{\mathrm{t}n},a_{\mathrm{w}n},b_{\mathrm{w}n})$ are the calibration coefficients, which are also functions of $x^*$ and $y^*$, as defined in equations \ref{trace} and \ref{wlsolution} (see also Table \ref{tab:coefficients}). 

\begin{figure}
	\centering
	\includegraphics[width=0.6\columnwidth]{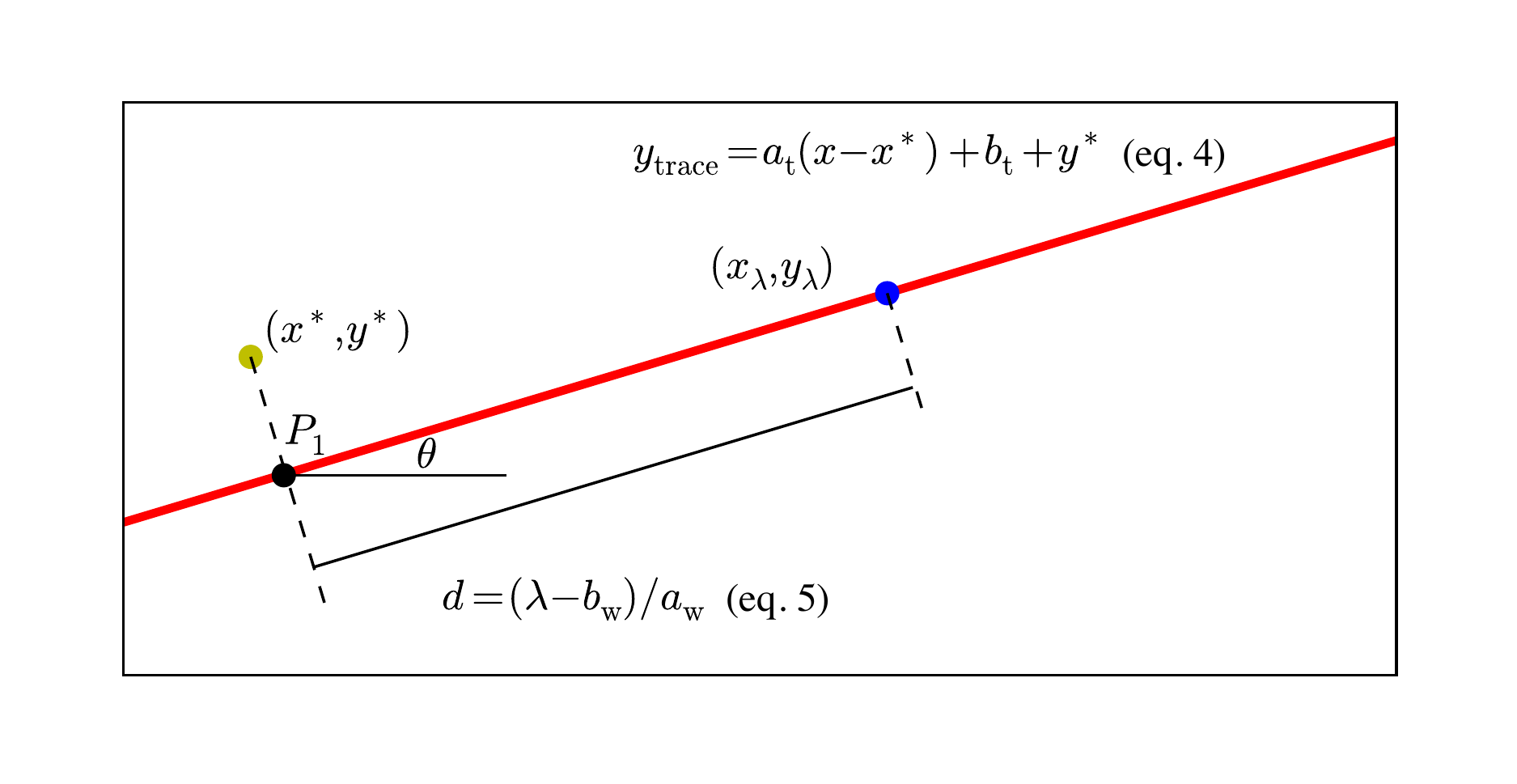}
	\caption{Relative positions of the trace (red line), the direct image of the star ($x^*,y^*$),  and a random point on the trace ($x_\lambda,y_\lambda$).}
	\label{fig:geometry}
\end{figure}

\begin{table}
	\small
	\centering
	\caption{Trace and wavelength solution calibration coefficients for the G141 grism \citep{coefficients}}
	\label{tab:coefficients}
	\begin{tabular}{c | c c c c c c}
		\hline \hline
						& $n=0$			& $n=1 \ [x^*]$		& $n=2 \ [y^*]$		& $n=3 \ [x^*y^*]$	& $n=4 \ [x^{*2}]$	& $n=5 \ [y^{*2}]$	\\ [0.1ex]
		\hline
		$a_{\mathrm{t}n}$	& 1.04275E-02		& -7.96978E-06		& -2.49607E-06		& 1.45963E-09		& 1.39757E-08		& 4.8494E-10		\\
		$b_{\mathrm{t}n}$	& 1.96882E+00		& 9.09159E-05		& -1.93260E-03		& 				&				&				\\
		$a_{\mathrm{w}n}$	& 4.51423E+01		& 3.17239E-04		& 2.17055E-03		& -7.42504E-07		& 3.48639E-07		& 3.09213E-07		\\	
		$b_{\mathrm{w}n}$	& 8.95431E+03		& 9.35925E-02		&				&				&				&				\\
	\end{tabular}
\end{table}

We derive them based on equations \ref{trace} and \ref{wlsolution}, and figure \ref{fig:geometry}, which shows the potion of the star, $P^*(x^*,y^*)$, and the photons of a particular wavelength, $P_\lambda (x_\lambda,y_\lambda)$, with respect to the spectrum trace (red line). Let $P_1(x_1,y_1)$ be the projection of $P^*$ on the trace. Because $P_1$ is on the trace, from equation \ref{trace} we have:
\begin{equation}
	y_1 -y^* = a_\mathrm{t} (x_1 - x^*) + b_\mathrm{t}
	\label{y1}
\end{equation}

\noindent Since $P_1$ is the projection of $P^*$ on the trace, the vectors $\overrightarrow{P^*P_1} = (x_1-x^*, y_1-y^*)$ and $\overrightarrow{V} = (1, a_t)$ (vector parallel to the trace) are orthogonal, so:
\begin{equation}
	\begin{split}
		\overrightarrow{P^*P_1} \cdot \overrightarrow{V} = 0 \Rightarrow x_1 - x^* + a_\mathrm{t}(y_1-y^*) = 0 \stackrel{\ref{y1}}{\Rightarrow} \\
		x_1 - x^* + a_\mathrm{t}(a_\mathrm{t} (x_1 - x^*) + b_\mathrm{t}) = 0 \Rightarrow (x_1 - x^*)(1 + a_\mathrm{t}^2) + a_\mathrm{t}b_\mathrm{t} = 0 \Rightarrow \\
		x_1 = x^* - \frac{a_\mathrm{t}b_\mathrm{t}}{1+a_\mathrm{t}^2}
	\end{split}
	\label{x1}
\end{equation}

\noindent Also, from Equation \ref{wlsolution}, the distance between $P_1$ and $P_\lambda$ along the spectrum trace is:
\begin{equation}
	d = \frac{ \lambda - b_\mathrm{w} }{ a_\mathrm{w} }
	\label{d}
\end{equation}

\noindent Finally, let $\theta$ be the inclination of the trace ($\theta$ = $\tan^{-1} (a_\mathrm{t})$) with respect to the x-axis of the detector:
\begin{equation*}
	\begin{split}
		\cos(\theta) = \frac{x_\lambda - x_1}{d} \Rightarrow x_\lambda = x_1 + d \cos [ \tan^{-1} (a_\mathrm{t}) ] \stackrel{\ref{x1},\ \ref{d}}{\Rightarrow} \\
		x_\lambda = x^* - \frac{a_\mathrm{t} b_\mathrm{t}}{1 + a_\mathrm{t}^2} + \frac{\lambda-b_\mathrm{w}}{a_\mathrm{w}} \cos [ \tan^{-1} (a_\mathrm{t}) ]
	\end{split}
\end{equation*}

\noindent $P_\lambda$ is also on the trace, so, from equation \ref{trace}:
\begin{equation*}
	y_{\lambda} = a_\mathrm{t}(x_{\lambda} - x^*) + b_\mathrm{t} + y^*
\end{equation*}
%Appendix: WAVELENGTH GRID EQUATIONS

%REFERENCES
{\small
\bibliographystyle{apj}
\bibliography{references} 
}
%REFERENCES

\end{document}